\documentclass[pre,twocolumn,twoside,byrevtex,superscriptaddress,floatfix]{revtex4-1}

\usepackage{settings}

\setboolean{twocolswitch}{true}
\graphicspath{ {./figures/} }

\begin{document}

\title{\protect %Socioeconomic relativity: A spatial study of communities \\ and their %nonlocal impact on health risks and mortality modeling 
% mThe socioeconomic relativity of death: Analyzing effects of\\geospatially-distributed variables in a Bayesian mortality model for Hong Kong
The sociospatial factors of death: Analyzing effects of geospatially-distributed\\variables in a Bayesian mortality model for Hong Kong}
%%%%%%%%%%%%%%%%%%%%%%%%%%%%%%%%%%%%%%%%%%%%%
\author{
\firstname{Thayer}
\surname{Alshaabi}
}
\email{thayer.alshaabi@uvm.edu}
\affiliation{ 
  Vermont Complex Systems Center,
  University of Vermont,
  Burlington, VT 05405.
} 
\affiliation{ 
  Computational Story Lab,
  University of Vermont,
  Burlington, VT 05405.
} 
\affiliation{
  Department of Computer Science,
  University of Vermont,
  Burlington, VT 05405.
  }

\author{
    \firstname{David Rushing}
    \surname{Dewhurst}
}
\affiliation{ 
  Vermont Complex Systems Center,
  University of Vermont,
  Burlington, VT 05405.
} 
\affiliation{ 
  Computational Story Lab,
  University of Vermont,
  Burlington, VT 05405.
} 
\affiliation{
  Charles River Analytics, 
  Cambridge, MA 02138.
  }

\author{
  \firstname{James P.}
  \surname{Bagrow}
}
%\email{james.bagrow@uvm.edu}
\affiliation{ 
  Vermont Complex Systems Center,
  University of Vermont,
  Burlington, VT 05405.
} 
\affiliation{
  Department of Mathematics \& Statistics,
  University of Vermont,
  Burlington, VT 05405.
  } 
  
\author{
  \firstname{Peter Sheridan}
  \surname{Dodds}
}
%\email{peter.dodds@uvm.edu}
\affiliation{ 
  Vermont Complex Systems Center,
  University of Vermont,
  Burlington, VT 05405.
} 
\affiliation{ 
  Computational Story Lab,
  University of Vermont,
  Burlington, VT 05405.
} 
\affiliation{
  Department of Computer Science,
  University of Vermont,
  Burlington, VT 05405.
  }
  
 \author{
  \firstname{Christopher M.}
  \surname{Danforth}
}
\email{chris.danforth@uvm.edu}
\affiliation{ 
  Vermont Complex Systems Center,
  University of Vermont,
  Burlington, VT 05405.
} 
\affiliation{ 
  Computational Story Lab,
  University of Vermont,
  Burlington, VT 05405.
} 
\affiliation{
  Department of Mathematics \& Statistics,
  University of Vermont,
  Burlington, VT 05405.
  }
%%%%%%%%%%%%%%%%%%%%%%%%%%%%%%%%%%%%%%%%%%%%%

\date{\today}

\begin{abstract}
  \protect

Human mortality is in part a function of multiple socioeconomic factors that differ both spatially and temporally.
Adjusting for other covariates, the human lifespan is positively associated with household wealth.
However, the extent to which mortality in a geographical region is a function of socioeconomic factors in both that region and its neighbors is unclear.
There is also little information on the temporal components of this relationship.
Using the districts of Hong Kong over multiple census years as a case study, 
we demonstrate that there are differences in how wealth indicator variables are associated with longevity in (a) areas that are affluent but neighbored by socially deprived districts 
versus (b) wealthy areas surrounded by similarly wealthy districts.
We also show that the inclusion of spatially-distributed variables reduces uncertainty in mortality rate predictions in each census year when compared with a baseline model. 
Our results suggest that geographic mortality models should incorporate nonlocal information (e.g., spatial neighbors) to lower the variance of their mortality estimates,
and point to a more in-depth analysis of sociospatial spillover effects on mortality rates.

\end{abstract}


%% 89.65.-s	Social and economic systems
%% 89.75.Da	Systems obeying scaling laws
%% 89.75.Fb	Structures and organization in complex systems
%% 89.75.-k	Complex systems (for complex chemical systems, see 82.40.Qt; for biological complexity, see 87.18.-h)

\maketitle

%%%%%%%%% end of author(s), address(es) plus abstract

%% add sections here...
%%%%%%%%%%%%%%%%%%%%%%%%%%%%%%%%%%%%%%%%%%%%%%%%%%%%%%%%%%%
\section{Introduction}\label{sec:introduction} 
Although Hong Kong is a small island territory, it exhibits significant variation in occupations, income, foreign inhabitant density, and residence status of workers.
In this study, we examine the benefits and drawbacks of incorporating nonlocal and spatial information into a mortality model for a limited area with restricted publicly available data.
Simulating a realistic scenario with limited spatial resolution, we show heterogeneity of such exogenous factors and investigate nonlocal behavioral interactions of prosperity and deprivation across neighborhoods.

We present an analytical evaluation comparing local and nonlocal models to show the importance of spatial associations for mortality modeling. 
In particular, we apply a spatial network technique to examine socioeconomic nonlocality among communities.
For instance, we investigate how the magnitude of a socially deprived area can consequently have a nonlocal effect on its neighbors’ mortality risks. 
Similarly, we delve into how the spatial spread of property of an affluent area can spillover to its surrounding areas, and thus affect their longevity. 
Our work not only reveals the deep influence of these spatial interactions of districts on predicting fatality rates, but also provides a method for investigating systematic inference errors of mortality models.

We structure our paper as follows. 
We discuss key findings of mortality risk studies in the literature and how they relate to our case examination in the next section. 
We introduce and analyze our data sources in Sec.~\ref{sec:data}.
We summarize the economic and social indicators used in our investigation in Appendix~\ref{sec:data_variables}.  
For our analytical inquiry, we employ a set of Bayesian generalized additive models to predict mortality rates across districts in Hong Kong. 
We describe our experiments and our exposition of the models in Sec.~\ref{sec:models}.
First, we present our local model that does not use any spatial information in Sec.~\ref{sec:local}. 
We compare our Baseline design to two nonlocal spatial models in Sec.~\ref{sec:nonlocal}. 
Our first nonlocal model uses spatial features from the nearest neighbours, while the second uses features from all neighbours weighted by their distance to the target area. 
We will refer to the nonlocal models as SP, and WSP respectively. 
We show our findings in Sec.~\ref{sec:results}, highlighting the computational complexity of each method and discussing the benefits and shortcomings of each design.
Our evaluation also reveals evidence of sociospatial spillovers of mortality rates. 
We conclude with some remarks on the limitations of our investigation and potential future work.

\section{Related work}\label{sec:background} 

\subsection{Mortality risks and social deprivation}

There are many studies that delve into the temporal dynamics of mortality risks with respect to nation-wide epidemics~\cite{wong2015breast,wu2017joint}, pollution~\cite{wong2001effect,qiu2015air}, and life expectancy~\cite{lam2004leisure} over the last decade.
Researchers have hypothesized and identified several connections of longevity, social deprivation, and socioeconomic discrimination~\cite{hayward1997inequality,deaton2003mortality,rey2009ecological}.
Notably, there are many interpretations of social deprivation. 
Messer~\etal’s~study~\cite{messer2006development} offers a well-written overview of socioeconomic deprivation in the literature. 
The authors highlight the limitations of such definitions and propose an alternative method to calculate and standardize what they call a ``neighborhood deprivation index’’ (NDI). 
Employing principal components analysis (PCA) on census data from 1995 to 2001, they illustrate the effectiveness of their proposed measurement at capturing socioeconomically deprived counties in the US.

Others have investigated a wide range of socioeconomic, psychological, and behavioral factors of fatality risks~\cite{puterman2020predicting}. 
We often examine the notion of disparity in health and mortality risks using population-scale inputs and sensitive individual variables such as age, race, and gender respecting the privacy concerns that emerge from such applications~\cite{santos2020differential}.
Ou~\etal~\cite{ou2008socioeconomic} infer socioeconomic status by type of housing, education, and occupation.
They find that regions with lower socioeconomic status have higher rates of air pollution.
They also report that neighborhoods with higher densities of blue-collar workers have higher rates of air pollution-associated fatality than others.
Chung~\etal~\cite{chung2018socioeconomic} present evidence of inequalities conditioned on age as a control variable.
The authors investigate the impact of socioeconomic status amid the rapid economic development of Hong Kong. 
Their findings suggest a decline in socioeconomic disparity in mortality risks across the distrcits of Hong Kong from 1976 to 2010. 
They also show that various health benefits brought by economic growth are greater for regions with higher socioeconomic status.
The market share of health benefits is unequally distributed among groups of varying status: Individuals with higher socioeconomic status have access to greater benefits than those of lower socioeconomic status.
In the present study, we use a set of socioeconomic attributes including income, unemployment, and mobility, to define and capture the some of the ramifications of social deprivation in Hong Kong.

\subsection{Spatial association of mortality risks}

Spatial associations between income disparity and health risks are widely understood both internationally, and for individual cities and states~\cite{buckingham1997sociodemographic,kim2008inequality,bjornstrom2011examination,chen2012role,fan2016tract,kim2018peer,yang2019modeling}. 
Local attributes play a powerful role in the model dynamics, given the assumption that socioeconomic factors vary geographically. 
Studies have shown the importance of spatial associations in identifying relations between socioeconomic deprivation and longevity. 
Although researches have examined income inequality, they often use a spatially localized approach in their investigations~\cite{kawachi1997relationship,ross2000relation,major2010neighborhood,reardon2011income,yang2015exploring}. 

Geographically weighted regression (GWR) is a commonly used method designed to examine spatial associations~\cite{fotheringham1998geographically, fotheringham2003geographically}.
Fotheringham~\etal argue socioeconomic features are intrinsically intra-connected over space because of the mechanisms by which communities develop. 
Their study makes an empirical comparison of their proposed method (GWR) to other stationary regression models to investigate the spatial distribution of long-term illness in the UK.

Others have looked into the spatial association between air pollution and mortality in Hong Kong~\cite{wong2008effects}, Czechia~\cite{branis2012association}, Rome~\cite{forastiere2007socioeconomic}, and France~\cite{padilla2014air}. 
Cossman~\etal~\cite{cossman2007persistent} examine the spatial distribution of mortality rates over 35 years, starting from 1968 to 2002 across all counties in the US. 
The study highlights a nonrandom pattern of clustering in mortality rates in the US, where high fatality rates are primarily driven by economic decline.

To assess geospatial associations between pollution and mortality in Hong Kong, Thach~\etal~\cite{thach2015assessing} examine the spatial interactions of tertiary planning units (TPUs)~\cite{tpu}---similar to census-blocks in the US.
The authors show a positive spatial correlation between mortality rates and seasonal thermal changes in Hong Kong. 
They argue that the variation between TPUs is a key factor for cause-specific fatality rates.  
Their results show that socioeconomically deprived regions have higher fatality rates, especially during winter.

\subsection{Sociospatial factors of death}

Studying the relative spatial interactions of social and economic indicators dates back to decades ago.
Researchers delve into measuring nonlocal and/or interdependent interactions of inequality in life expectancy~\cite{boing2020quantifying}, health care~\cite{erreygers2011measuring}, 
education~\cite{zimmer2000peer,sacerdote2011peer}, and decision-making~\cite{bodine2013conforming,albert2013teenage}.
Many methods have been proposed to identify and examine broader dimensions of inequality from a spatial point of view 
such as Moran’s I and spatial auto-regression~\cite{moran1950notes,li2007beyond,yang2011social}.
Yang~\etal~\cite{yang2015exploringdurbin} argue that mortality rates of counties in the US 
are associated with social and economical aspects found in neighboring counties. 
Their findings suggest that fatality rates in a county are remarkably driven 
by social signals from bordering counties 
because of the spillover of socioeconomic wealth or social deprivation across neighborhoods. 
Another recent work by Holtz~\etal~\cite{holtz2020interdependence} 
highlights the significant influence of nonlocal interactions and spillovers 
on regional policies regarding the global outbreak of COVID--19.
Employing a network-based approach to explore the dynamics of communities and their impact on mortality risks, 
we present here a small-case study using a collection of datasets from Hong Kong. 
In our study, we use three different models to illustrate the role of spatial associations 
by comparing models with spatial features to a baseline model without spatial factors.

\section{Methods}\label{sec:methods}
\subsection{Data sources}
\label{sec:data}

\begin{figure*}[tp!]
\centering 
\includegraphics[width=\textwidth]{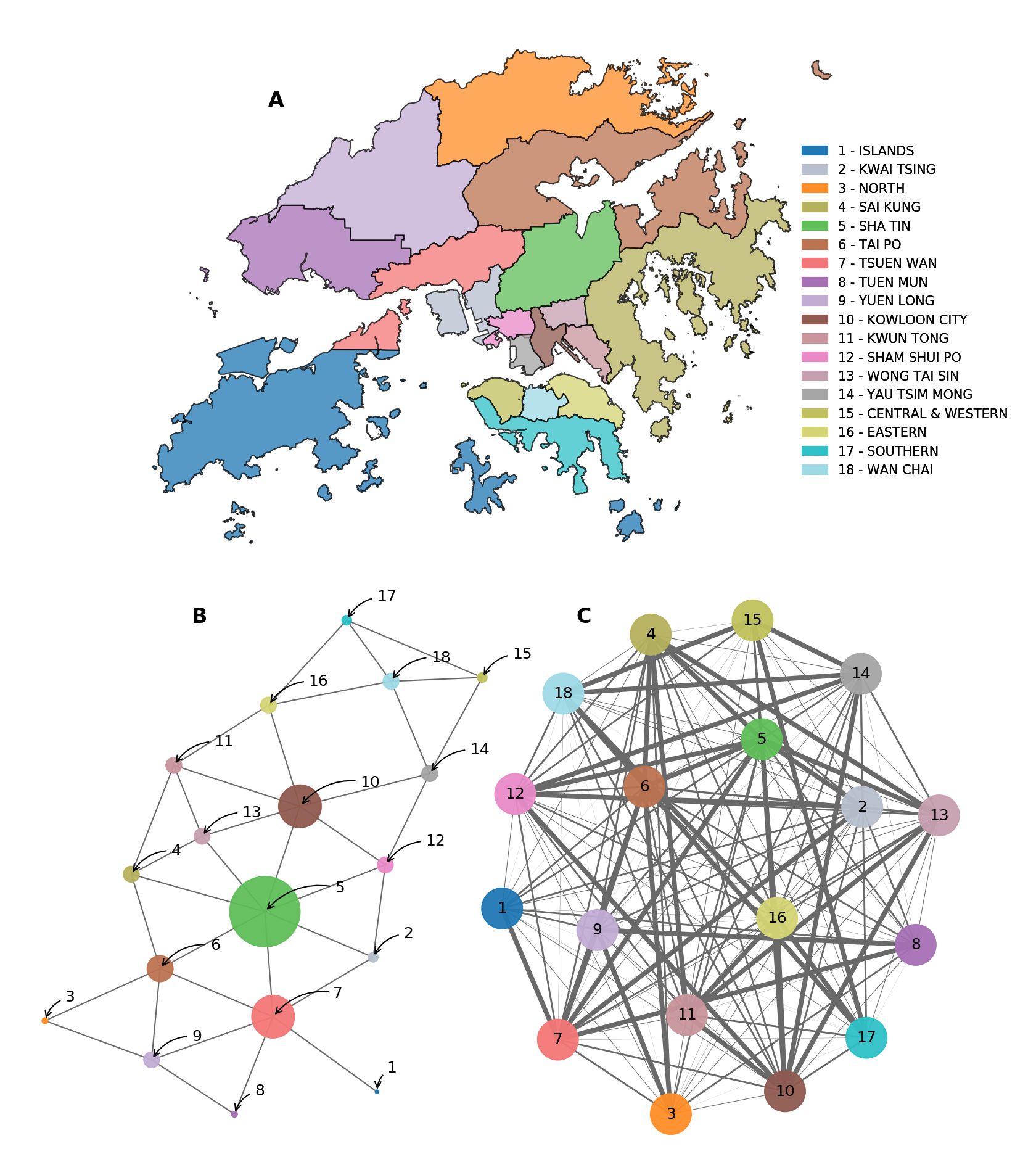}
\caption{
\textbf{Spatial networks of Hong Kong's districts}.
We cross-reference the roads and bridges connecting these districts to build a spatial network~\cite{zhou2016empirical,chen2012vulnerability}. 
(\textbf{B}) We demonstrate the first undirected network layout of Hong Kong's districts. 
Districts (nodes) are linked if they border each other or share a direct road/bridge in a binary fashion.
(\textbf{C}) We show a fully connected version of the network. 
For the fully connected network, edges to neighboring districts are weighted exponentially. 
Different weighting schemes can be applied here, 
however, for our application we use the spatial distance measured by the length of the shortest path connecting any two districts on the network. 
}
\label{fig:network}
\end{figure*} 

\paragraph*{Census data:} 
We collected socioeconomic variables curated by the 
Census and Statistics Department of Hong Kong~\cite{hongkong_census_dataset}.
We have three snapshots at 5-year intervals, 2006, 2011, and 2016.
For each year, the dataset includes the total population density by district, 
median income, 
median rent to income ratio,
median monthly household income,
unemployment rates, 
unemployment rates across households, 
unemployment rates among minorities,
the proportion of homeless people,
the proportion of homeless mobile residents,
the proportion of single parents,
the proportion of households with children in school,
the proportion of households with children (aged under 15),
and the proportion of households with elderly (aged 65 or above).

\paragraph*{Mortality data:}
We use the official counts of 
known and registered deaths provided by the Census and Statistics Department of Hong Kong~\cite{hongkong_mortality_dataset}. 
The data set contains 892,055 death records between 1995 and 2017.
Every record includes a wide range of information such as age, gender,
and place of residence 
(TPU)~\cite{tpu}---a geospatial reference system used
to report population census statistics.
Our mortality records have both \textit{place of occurrence} and \textit{place of residence}. 
Each of those spatial features is certainly important 
to paint a better picture of the microscopic 
spatial associations of mortality and other social and economic factors. 
For our study, however, 
we only use \textit{place of residence} as our primary geographic unit 
and discard all years other than the census years
to cross reference our death records with other socioeconomic characteristics for each district. 
We derive an annual crude death rate of each district by simply dividing the number of registered death records by the population size for each calendar year such that: 
$$Y_{it} = \dfrac{N_{it}(\textnormal{deaths})}{N_{it}(\textnormal{census population})},$$
for district $i$ and $t \in [2006, 2011, 2016]$.

\paragraph*{Life insurance data:}
We have obtained data from a Hong Kong based life insurance provider.
According to the Hong Kong Insurance Authority~\cite{HKIA}, our provider had roughly 2.5\% market share of all non-linked individual life insurance policies issued in Hong Kong in 2016.
We normalize the number of polices issued at the district level by population size for each time snapshot to report the proportion of individuals insured by each district. 
Notably, our variable is limited to policies sold by a single company and thus affected by the sociospatial features of the company market share such as the spatial sparsity of its agents and offices, and the social characteristics of consumers who would choose our provider over other life insurance providers in the area.  
However, statistical and detailed data sources regarding life insurance policies are often proprietary, especially with a similar spatial resolution to the one presented in our study.  
Although our data on life insurance policies may not represent the full population, it provides an example of the data that an insurance company can use to build their models.
Given the scarcity of such data, we use our records of life insurance policies as a useful complementary wealth indicator---among other variables such as income, rent and unemployment---which is absent from most studies.

\paragraph*{Geospatial unit:}
Initially, we planned to use TPUs as the main geospatial units to cross-reference our data sources. 
However, we identified a large subset of missing TPUs in death records, as records in small TPUs may reveal sensitive information about specific individuals there.
To avoid any risk of identifying individuals in the data set, we use districts as our main spatial unit of analysis~\cite{districts}.
This choice is consistent with prior work, where most studies have either filtered out small TPUs in their analyses~\cite{wong2008effects,chung2018socioeconomic}, or aggregated their records at the district level~\cite{thach2015assessing} to overcome this challenge.

\paragraph*{Categorization:}
We organize our features into three different categories.

\begin{enumerate}
    \item \textbf{Base}: This set has most of the socioeconomic features in our data sets such as population density, unemployment rates, the proportion of homeless people, mobile residents, and single-parent households. 
    However, we do not include wealth-, age-, or race-related features here.
    \item \textbf{Wealth}: Besides the base features described above, we include median income, median rent-to-income ratio, median monthly household income, and life insurance coverage by the district. 
    \item \textbf{All}: This set includes all features in our data set, including sensitive variables---from a sociopolitical perspective---such as the proportion of minorities and unemployment rates among minorities. This set also includes age-related features such as the proportion of young and elderly residents for each district. 
\end{enumerate}

\subsection{Statistical methods}
\label{sec:models}

There are many statistical modeling paradigms to tackle this task.
Each of which comes with its own costs and benefits.   
Researchers sometimes use Poisson models to investigate mortality risks~\cite{wong2008effects,qiu2015air,thach2015assessing,wong2015breast}.
Others use general linear or multivariate regression models~\cite{kiffer2011spatial,padilla2014air,boing2020quantifying}.
We can also see modifications of this family of approaches in the literature such as geographically weighted regression~\cite{fotheringham1998geographically,fotheringham2003geographically}.
In this study, however, we consider the simplest approach.
We use a set of three Bayesian multivariate linear regression models.
Our goal is to examine the addition of nonlocal information, regardless of the distributional assumptions placed on the response variable $Y$. 
Therefore, we keep our model as simple as possible to allow us to investigate the implications of two different spatial models compared to a baseline model that does not factor in any nonlocal information.
We treat the design tensors $\vec X$ as exogenous variables and do not model their evolution across time.
A ``design tensor’’ is a rank 3 tensor given by $\vec X = (\vec X_1, \vec X_2, \ldots, \vec X_T)$ where $\vec X_t$ is the design matrix for time period $t \in [2006, 2011, 2016]$.
Each design matrix $\vec X_t$ is of dimension $N \times (p + 1)$, where $N$ is the number of observations which accounts for 18 districts in Hong Kong, and $p$ is the number of predictors.
We add an extra variable to the design matrix to account for a constant in our linear model.

\subsubsection{Local model (Baseline)}\label{sec:local}

The dynamics of the local models are described by a system of linear equations,
\begin{align}
    \vec y_t 
    &= 
    \vec X_t \vec \beta_t + \sigma_t \vec u_t, \label{eq:static-lm} \\
    \vec \beta_t
    &=
    \vec \beta_{t-1} + \vec \mu + \vec L
    \vec v_{t}, \label{eq:beta-rw} \\
    \log \sigma_t 
    &= 
    \log \sigma_{t - 1} + \mu + \ell w_t,
    \label{eq:log-sigma-rw}
\end{align}
for $t = 1,...,T$.

Eq.~\ref{eq:static-lm} is an ordinary linear model for the response vector $\vec y_t$
as a function of the design matrix $\vec X_t$ and coefficients $\vec \beta_t$.
We presently define the quantities that compose Eqs.\ \ref{eq:beta-rw} and \ref{eq:log-sigma-rw}.
We set 
\begin{equation}
    \vec u_t, \vec v_t \sim 
\text{MultivariateNormal}(\vec 0,\vec I)
\end{equation}
in Eqs.~\ref{eq:static-lm} and \ref{eq:beta-rw}, while
$w_t \sim \text{Normal}(0, 1)$. 
Our identity matrix $\vec I$ is informed by the number of predictors in our model,
and has a dimension of $(p + 1) \times (p + 1)$. 
Hence the model likelihood is  
\begin{equation}
    \begin{aligned}
    p(\vec y | \vec \beta, \sigma) &= \prod_{t=1}^T\prod_{n=1}^N 
    p(y_{tn} | \vec X_{tn} \vec \beta_t, \sigma_t) \label{eq:likelihood}\\
    &= \prod_{t=1}^T\prod_{n=1}^N \text{Normal}(\vec X_{tn} \vec \beta_t, \sigma_t^2).
    \end{aligned}
\end{equation}
\begin{figure}[tp!]
\centering
\includegraphics[width=\columnwidth]{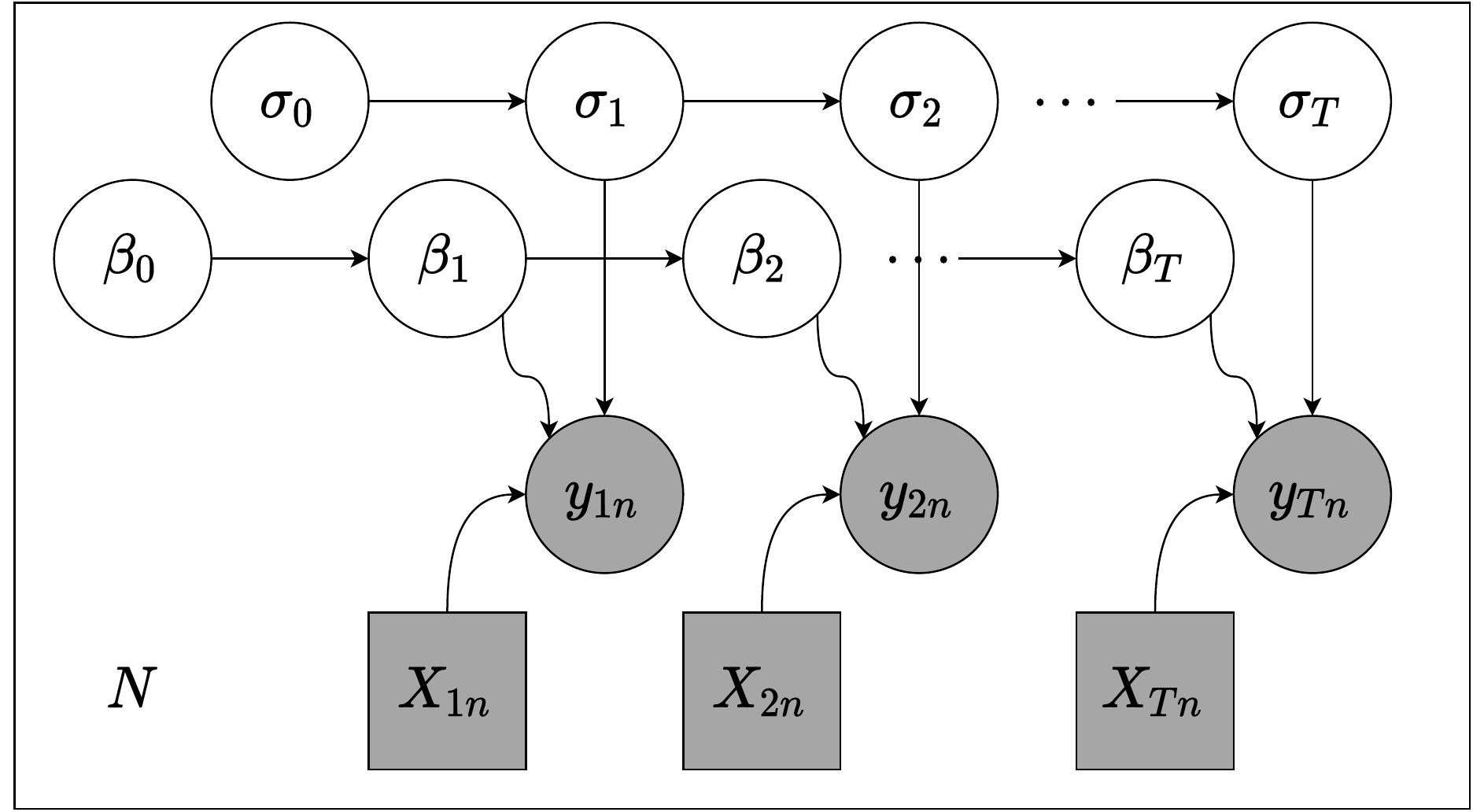}
\caption{
\textbf{A graphical model representing the likelihood function given in Eq.~\ref{eq:likelihood}.}
Latent $\log \sigma_t$ and $\beta_t$ evolve as biased random walks, 
while $y_{tn}$ and $\vec X_{tn}$ are treated as observable random variables and exogenous parameters respectively.
The entire temporal model is plated across districts $N = 18$.
}
\label{fig:dag-likelihood}
\end{figure}

A graphical model corresponding to Eq.~\ref{eq:likelihood} is displayed in Fig.~\ref{fig:dag-likelihood}.
We also \textit{a priori} believe that $\vec \beta_t$ does not remain constant throughout the time under study, 
though we are unsure of exactly how it changes over this time.
Thus, we assume a prior on $\vec \beta_t$ that evolves as a biased random walk with drift 
given by $\vec \mu$ and correlation matrix 
$\vec \Sigma$ with Cholesky decomposition $\vec L$.

Likewise, we suppose that $\log \sigma_t$ evolves according to a univariate random walk
with drift 
given by $\mu$ and standard deviation $\ell$.
We make this assumption for the same reason: We do not believe it is likely that $\sigma_t$ 
remains constant over the entire time period of study.
The random walk priors for $\vec \beta_t$ and $\sigma_t$ are each centered about zero 
because we impose a zero mean prior on  $\vec \mu$ and $\mu$.
We initialize these random walks with zero-centered multivariate normal initial conditions, 
\begin{equation}
    \vec \beta_0 \sim 
    \text{MultivariateNormal}(\vec 0, \vec \Sigma),
\end{equation}
and 
\begin{equation}
    \log \sigma_0 \sim 
    \text{Normal}(0, \ell^2).
\end{equation}
The distribution of $\vec\beta_{1:T} \equiv (\vec\beta_1,...,\vec\beta_T)$ is thus given by
\begin{align}
    p(\vec \beta_{1:T}|\vec \mu, \vec \Sigma)
    = \prod_{t=1}^T 
    p(\vec \beta_t|\vec \beta_{t-1}, \vec \mu, \vec \Sigma)\\
    = \prod_{t=1}^T 
    \text{MultivariateNormal}(\vec \beta_{t-1} + \vec \mu, \vec \Sigma),
\end{align}
with an analogous (univariate) distribution holding for $\log \sigma$.
We set
\begin{align}
\vec \mu
&\sim
\text{MultivariateNormal}(\vec 0, \vec I),\\ \mu 
&\sim \text{Normal}(0, 1),
\end{align}
so that the prior distribution over the paths of the regression coefficients and log 
standard deviation are centered about zero---the null hypothesis---for all $t$.

We place a uniform prior (LKJ(1)) over the correlation component of $\vec \Sigma$. The mean of this prior is at the identity matrix.
The vector of standard deviations of $\vec \Sigma$, $\vec s$, is hypothesized to follow an isotropic multivariate 
log normal distribution as 
$\vec s \sim \text{LogNormal}(0, 1)$.
We also place a univariate $\text{LogNormal}(0, 1)$ prior on $\ell$, the standard deviation of the increments of $\log \sigma_t$. 
We make this choice because we do not possess prior information about the appropriate noise scale of $\vec \beta_t$ or $\log \sigma_t$ and the log normal distribution is a weakly-informative prior that does not encode much prior information about their noise scales.

We did not perform exact inference of this model but rather fit parameters of a surrogate variational posterior distribution.
We use here variational inference to approximate the posterior probability of the design tensor and ultimately run Bayesian inference over these features~\cite{beal2003variational,hoffman2013stochastic}. 
Although traditional methods such as Markov chain Monte Carlo sampling (MCMC) can offer guarantees of accurate sampling from the target density, it does not necessarily outperform variational inference in terms of accuracy~\cite{braun2010variational,ranganath2014black,kucukelbir2017automatic}.
Evaluating the costs and benefits for accurate estimation of posteriors is indeed an open area of research, however, variational inference offers a much faster and effective method to approximate probability densities through optimization, even for small datasets~\cite{kucukelbir2015automatic,blei2017variational}.
Using variational inference also allows us to have an agile development cycle and flexible models, as access to more economic and social data features will continue to evolve, and change over time. 
The effectiveness and versatility of variational inference has left a remarkable positive impact across disciplines~\cite{ghahramani2001propagation,rossi2012bayesian,barber2012bayesian,zhang2018advances,bottou2018optimization}.

Denoting the vector of all latent random variables by
\begin{align}
\vec z \equiv (
\mu^{(\sigma)},
\ell,
\vec \mu^{(\vec \beta)}, 
\Sigma^{(\vec \beta)},
\log \sigma_{0:T},
\vec \beta_{0:T} 
),
\end{align}
we fit the parameters $\psi$ of an approximate posterior distribution 
$q_{\psi}(\vec z)$ to maximize the variational lower bound, defined as the expectation under $q_\psi(\vec z)$ of the difference between the log joint probability and $\log q_\psi(\vec z)$~\cite{hoffman2013stochastic}.
We chose a low-rank multivariate normal guide with rank equal to approximately 
$\sqrt{\text{dim}\ \vec z}$.
This low rank approximation allows for modeling of correlations in the posterior distribution of $\vec z$ with a lower number of parameters than, for example, a full-rank multivariate normal guide distribution. 
All bounded latent random variables are reparameterized to lie in an unconstrained space so that we could approximate them with the multivariate normal guide. 

\begin{figure*}[!th]
\centering 
\includegraphics[width=\textwidth]{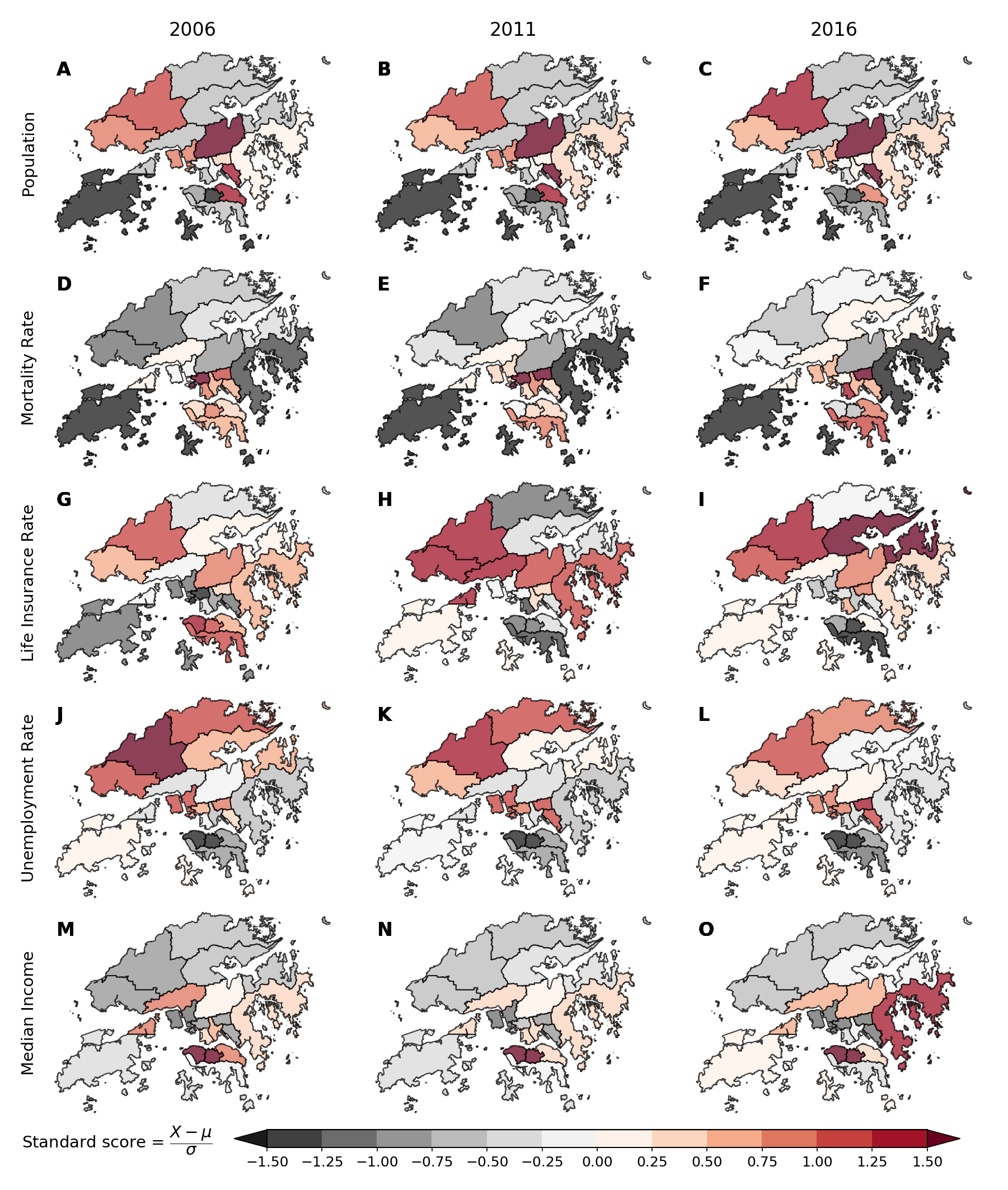}
\caption{
    \textbf{Temporal dynamics of spatial socioeconomic characteristics of Hong Kong}.
    We show the spatial distribution of five features in our datasets for three different census years. 
    Here, heatmaps are normalized by the mean and standard deviation.
    Darker shades of red show areas above the mean for each of these variables while shades of grey show areas below the mean.
    (\textbf{A--C}) We display the spatial growth of population over time. 
    (\textbf{D--F}) We demonstrate the variation of mortality rates, 
    and life insurance converge (\textbf{G--I}).
    We see some segregation of unemployment rates in (\textbf{J--L}),
    and median income in (\textbf{M--L}).
    }
\label{fig:map}
\end{figure*}

\subsubsection{Nonlocal models (SP and WSP)}\label{sec:nonlocal}

We use the road networks in Hong Kong to build a spatial network of the 18 districts~\cite{zhou2016empirical,chen2012vulnerability}. 
Each node in the network represents a single district. 
Nodes are linked if they share a direct road or bridge.
In Fig.~\ref{fig:network}A, we show a map of Hong Kong’s districts. 
We display an undirected network of districts in Fig.~\ref{fig:network}B.
By contrast, we show a fully connected version of the network in Fig.~\ref{fig:network}C. 
Edges are weighted by their spatial distance measured by the length of the shortest path $d_{ij}$ to reach from district $i$ to district $j$
\begin{align}
    w_{ij} = \text{exp}\{-(d_{ij}-1)\},
\end{align}
and weights decays exponentially as the length of the shortest path increases between any two districts.

Similarly, we treat the adjacency matrix $\vec A$ as an exogenous variable.
We fit two nonlocal models that leverage the design matrix associated with each district's neighbors;
a spatial model (SP) that uses the binary adjacency matrix and 
a weighted spatial model (WSP) that uses the weighted adjacency matrix.
The equations describing the time evolution of this data generating process are 
\begin{align}
    \vec y_t 
    &= 
    \vec X_t \vec \beta_t + 
    \vec f\left(\vec X_t \otimes \vec B \right) \vec \gamma _t + 
    \sigma_t \vec u_t, \label{eq:network-lm} \\
    \vec \beta_t
    &=
    \vec \beta_{t-1} + \vec \mu + \vec L \vec v_{t}, \label{eq:beta-network} \\
    \vec \gamma_t
    &=
    \vec \gamma_{t-1} + \vec m + \vec \Lambda \vec q_{t}, \label{eq:gamma-network} \\
    \log \sigma_t 
    &= 
    \log \sigma_{t - 1} + \mu + \ell w_t,
    \label{eq:log-sigma-network}
\end{align}
for $t = 1,...,T$.
The rank three tensor $\vec X_t \otimes \vec B$ is the outer product of
$\vec B \equiv \vec A - \vec I$
with the design matrix $\vec X_t$.
The function $\vec f$ is a reduction function that lowers the rank of the tensor by one by collapsing the first dimension. Here we take $\vec f$ to be the mean across the first dimension.
In other words, $\vec f\left(\vec X_t \otimes \vec B \right)$ is a design matrix where $\vec f\left(\vec X_t \otimes \vec B \right)_{ij}$ is the average of the values of predictor $j$ over all the neighbors of district $i$ in the network.

The prior distributions for $\vec \gamma_t$, $\vec m$, $\vec \Lambda$, and 
$\vec q_t$ are identical to those for $\vec \beta_t$, $\vec \mu$, $\vec L$,
and $\vec u_t$ except their dimensionality is lowered from $p + 1$ to $p$ since we do not include an intercept term in $\vec f\left(\vec X_t \otimes \vec B \right)$.

We use \texttt{Pyro}~\cite{bingham2018pyro}, a probabilistic programming language that operates on top of \texttt{Pytorch}~\cite{NEURIPS2019_9015_PyTorch}, a dynamic graph differentiable programming library, to implement our models. 
Our source code is available on Gitlab \footnote{https://gitlab.com/compstorylab/asis}.

\section{Results and Discussion}\label{sec:results} 

\subsection{Observational and analytical findings}

In Fig.~\ref{fig:map}, we display the spatial distribution of socioeconomic characteristics for 2006, 2011, and 2016. 
Each heatmap is normalized by the mean and standard deviation for each year, such that darker shades of red show areas above the mean for each of these variables, while shades of grey show areas below the mean.
We show normalized population density in Figs.~\ref{fig:map}A through C.
We see dense clusters both at the center of the country and on the northwestern side.

We are primarily interested in the geospatial trend across different variables/predictors for each year, respectively.
For example, our heatmaps in Figs.~\ref{fig:map}D--F show that the southern islands have higher mortality rates than the average rates of Hong Kong.
The southern islands had higher rates of new life insurance policies in 2006 (Fig.~\ref{fig:map}G), followed by consistently lower rates than average when compared to the rest of the districts in Hong Kong for each year, respectively (see Figs.~\ref{fig:map}H--I).

The northwestern territories have higher rates of unemployment compared to the southeastern side of Hong Kong, as we see in Figs.~\ref{fig:map}J--L.
In Figs.~\ref{fig:map}M--O, we observe that the east and center districts have higher normalized median income when adjusted for inflation.
We display additional statistics regarding households in Fig.~\ref{fig:household}.

\subsection{Evaluation and comparison of the models}

\begin{table*}[th!]
\centering
\caption{\textbf{Model evaluation}.
We report the mean absolute error for each model across all districts averaged over a 1000 trials.
The cells colored in blue show the model with the lowest margin of error for each feature category, whereas grey cells demonstrate ties among two models.
The model highlighted in red indicates the model with the lowest margin of error.
}
\def\arraystretch{1.5}
\begin{tabular}{l || x{13mm} x{13mm} x{13mm} | x{13mm} x{13mm} x{13mm} | x{13mm} x{13mm} x{13mm}}
\textbf{Mean absolute error} & \multicolumn{3}{c}{\textbf{Base} (7 predictors)} & \multicolumn{3}{c}{\textbf{Wealth} (11 predictors)} & \multicolumn{3}{c}{\textbf{All} (16 predictors)}\\
$e^t = \frac{1}{N} \sum_{i=1}^{18} |\hat{Y}^t_i - Y^t_i|$ & \footnotesize \textbf{Baseline} & \footnotesize \textbf{SP} & \footnotesize \textbf{\cellcolor{red!25}WSP} & \footnotesize \textbf{Baseline} & \footnotesize \textbf{\cellcolor{red!25}SP} & \footnotesize \textbf{WSP} & \footnotesize \textbf{\cellcolor{red!25}Baseline} & \footnotesize \textbf{SP} & \footnotesize \textbf{WSP} \\
\hline
2006 &     1.19 & 1.25 & \cellcolor{blue!25}1.17 &    1.25 & \cellcolor{blue!25}1.22 & 1.25 &     \cellcolor{blue!25}1.32 & 1.40 & 1.46 \\
2011 &     0.99 & 1.03 & \cellcolor{blue!25}0.94 &     1.02 & \cellcolor{blue!25}1.00 & 1.11 &     \cellcolor{blue!25}1.13 & 1.24 & 1.37 \\
2016 &     0.90 & 0.92 & \cellcolor{blue!25}0.86 &    \cellcolor{gray!25}1.09 & \cellcolor{gray!25}1.09 & 1.26 &     \cellcolor{blue!25}1.20 & 1.34 & 1.51 \\
\end{tabular}
\label{tab:errors_avg}
\end{table*}

\begin{table*}[th!]
\centering
\caption{\textbf{Model evaluation by district}.
We report the mean absolute error for each model in 2016 averaged over a 1000 trials.
The cells colored in blue show the model with the lowest margin of error for each district and feature category,
whereas grey cells demonstrate ties among two models for a district.
The model highlighted in red indicates the model with the lowest margin of error across all districts, respectively.
}
\def\arraystretch{1.5}
\begin{tabular}{l || x{13mm} x{13mm} x{13mm} | x{13mm} x{13mm} x{13mm} | x{13mm} x{13mm} x{13mm}}
\textbf{Mean absolute error}  & \multicolumn{3}{c}{\textbf{Base} (7 predictors)} & \multicolumn{3}{c}{\textbf{Wealth} (11 predictors)} & \multicolumn{3}{c}{\textbf{All} (16 predictors)}\\
$e_i^{2016} = |\hat{Y}_i^{2016} - Y_i^{2016}|$ & \footnotesize \textbf{Baseline} & \footnotesize \textbf{SP} & \footnotesize \textbf{\cellcolor{red!25}WSP} & \footnotesize \textbf{Baseline} & \footnotesize \textbf{\cellcolor{red!25}SP} & \footnotesize \textbf{WSP} & \footnotesize \textbf{\cellcolor{red!25}Baseline} & \footnotesize \textbf{SP} & \footnotesize \textbf{WSP} \\
\hline
ISLANDS           &     1.39 & \cellcolor{blue!25}1.03 & 2.09 &     \cellcolor{blue!25}1.71 & 2.37 & 2.07 &     \cellcolor{blue!25}2.09 & 2.72 & 3.17 \\
KWAI TSING        &     0.82 & 0.65 & \cellcolor{blue!25}0.61 &     0.78 & \cellcolor{blue!25}0.65 & 1.01 &     \cellcolor{gray!25}0.95 & \cellcolor{gray!25}0.95 & 1.18 \\
NORTH             &     0.59 & 0.84 & \cellcolor{blue!25}0.54 &     1.42 & \cellcolor{blue!25}1.04 & 1.39 &     1.93 & \cellcolor{blue!25}1.72 & 1.90 \\
SAI KUNG          &     \cellcolor{gray!25}1.41 & \cellcolor{gray!25}1.41 & 1.78 &     \cellcolor{blue!25}1.52 & 1.99 & 1.63 &     \cellcolor{blue!25}1.59 & 1.82 & 1.89 \\
SHA TIN           &     1.26 & 1.43 & \cellcolor{blue!25}1.15 &     \cellcolor{gray!25}1.40 & \cellcolor{gray!25}1.40 & 1.62 &     \cellcolor{blue!25}1.56 & 1.72 & 1.81 \\
TAI PO            &     \cellcolor{gray!25}0.62 & 0.88 & \cellcolor{gray!25}0.62 &     \cellcolor{blue!25}0.85 & 0.96 & 1.11 &     \cellcolor{blue!25}0.94 & 1.33 & 1.40 \\
TSUEN WAN         &     \cellcolor{blue!25}0.36 & 0.57 & 0.49 &     \cellcolor{gray!25}0.55 & \cellcolor{gray!25}0.55 & 0.67 &     \cellcolor{blue!25}0.63 & 0.78 & 0.77 \\
TUEN MUN          &     0.65 & 0.71 & \cellcolor{blue!25}0.60 &     1.47 & \cellcolor{blue!25}1.16 & 1.72 &     1.49 & \cellcolor{blue!25}1.18 & 1.45 \\
YUEN LONG         &     \cellcolor{gray!25}0.63 & 0.81 & \cellcolor{gray!25}0.63 &     \cellcolor{blue!25}0.54 & 0.68 & 0.80 &     \cellcolor{blue!25}0.69 & 1.00 & 0.99 \\
KOWLOON CITY      &     \cellcolor{blue!25}0.41 & 0.67 & 0.54 &     \cellcolor{blue!25}0.51 & 0.65 & 0.74 &     \cellcolor{blue!25}0.68 & 0.94 & 0.98 \\
KWUN TONG         &     \cellcolor{blue!25}0.61 & 0.77 & 0.83 &     0.96 & \cellcolor{blue!25}0.84 & 1.09 &     \cellcolor{blue!25}1.15 & 1.47 & 1.46 \\
SHAM SHUI PO      &     0.55 & 0.46 & \cellcolor{blue!25}0.38 &     0.60 & \cellcolor{blue!25}0.52 & 0.77 &     \cellcolor{gray!25}0.73 & \cellcolor{gray!25}0.73 & \cellcolor{gray!25}0.73 \\
WONG TAI SIN      &     1.79 & 1.80 & \cellcolor{blue!25}1.70 &     2.29 & \cellcolor{blue!25}1.95 & 2.47 &     1.72 & \cellcolor{blue!25}1.25 & 1.45 \\
YAU TSIM MONG     &     0.79 & \cellcolor{blue!25}0.51 & 0.71 &     1.37 & \cellcolor{blue!25}1.19 & 1.22 &     \cellcolor{blue!25}0.81 & 0.96 & 0.92 \\
CENTRAL \& WESTERN &    \cellcolor{gray!25}0.75 & 0.82 & \cellcolor{gray!25}0.75 &     \cellcolor{gray!25}1.15 & \cellcolor{gray!25}1.15 & 1.61 &     \cellcolor{blue!25}1.91 & 2.20 & 2.95 \\
EASTERN           &     0.65 & \cellcolor{blue!25}0.56 & 0.77 &     0.78 & \cellcolor{blue!25}0.75 & 0.94 &     \cellcolor{blue!25}0.70 & 0.96 & 0.97 \\
SOUTHERN          &     1.42 & 1.80 & \cellcolor{blue!25}1.23 &     1.06 & \cellcolor{blue!25}0.92 & 1.32 &     \cellcolor{blue!25}1.34 & 1.41 & 1.85 \\
WAN CHAI          &     \cellcolor{gray!25}0.77 & 0.95 & \cellcolor{gray!25}0.77 &     \cellcolor{blue!25}0.60 & 0.79 & 0.79 &     \cellcolor{blue!25}0.69 & 1.03 & 1.12 \\
\end{tabular}
\label{tab:errors}
\end{table*}

Although we have similar and simple building blocks for our models, they do scale differently in terms of their computational costs. 
The total number of parameters in our Baseline model is equal to $pN + 1$, where $p$ is the number of predictors we use for each district. 
Besides the set of predictors for each target district, our spatial model SP uses spatial features from the nearest neighbours (ego-network) to that district. 
Thus, the total number of parameters used in the SP model is $\propto pN\bar{C}$ where $\bar{C}$ is the average clustering coefficient of the network.
Our WSP model leverages features from all neighbours weighted by their distance to the target district.
It has the largest number of parameters, which is proportional to $pN^2$.
The relative difference in the number of parameters among these models urges us to further investigate the benefits of expanding the models with spatial features.

To evaluate our models, we consider two metrics: mean absolute error (MAE) to estimate our margin of error and mean signed deviation (MSD) to examine systematic bias. 
In Table~\ref{tab:errors_avg}, we report the mean absolute error defined such that: 
$$e^t = \frac{1}{N} \sum_{i=1}^{N=18} |\hat{Y}^t_i - Y^t_i|,$$ 
for each calendar year $t \in [2006, 2011, 2016]$ in our dataset across all districts. 
We highlight cells in blue to show the model with the lowest margin of error, and red to indicate the best model for all years. 
We also color cells in grey to demonstrate a tie between two models for a year.

For our default set of features (Base), we note our WSP model outperformed the rest of the models in most districts. 
However, as we add more predictors to the models, we observe a pattern whereby models with fewer parameters perform better.
Our results show the SP model has the lowest MAE across districts when we use the Wealth category, which has 11 predictors including some wealth-related features as described in Sec.~\ref{sec:data}. 
The Baseline model---which does not account for spatial information---has a lower MAE when we feed all predictors to the models. 
This is an expected behavior because our larger and richer spatial models get overwhelmed with too many parameters and very limited data points.
We show a detailed breakdown for each model and each district for the calendar year 2016 in Table~\ref{tab:errors}.

We also compute a probability density function (PDF) of signed deviation ($\hat{Y}^t_i - Y^t_i$) for each model, which is possible since our models are fully Bayesian and generate a distribution of possible outcomes.
If the models accurately associate features with observed mortality rate, the distributions would be centered on zero. 
Conversely, if the models display systematic over- or under-estimation of mortality rate, the distributions will diverge away from zero, whereby negative numbers show underestimation and positive numbers indicate overestimation.

\begin{figure*}[th!]
\centering
\includegraphics[width=1.01\textwidth]{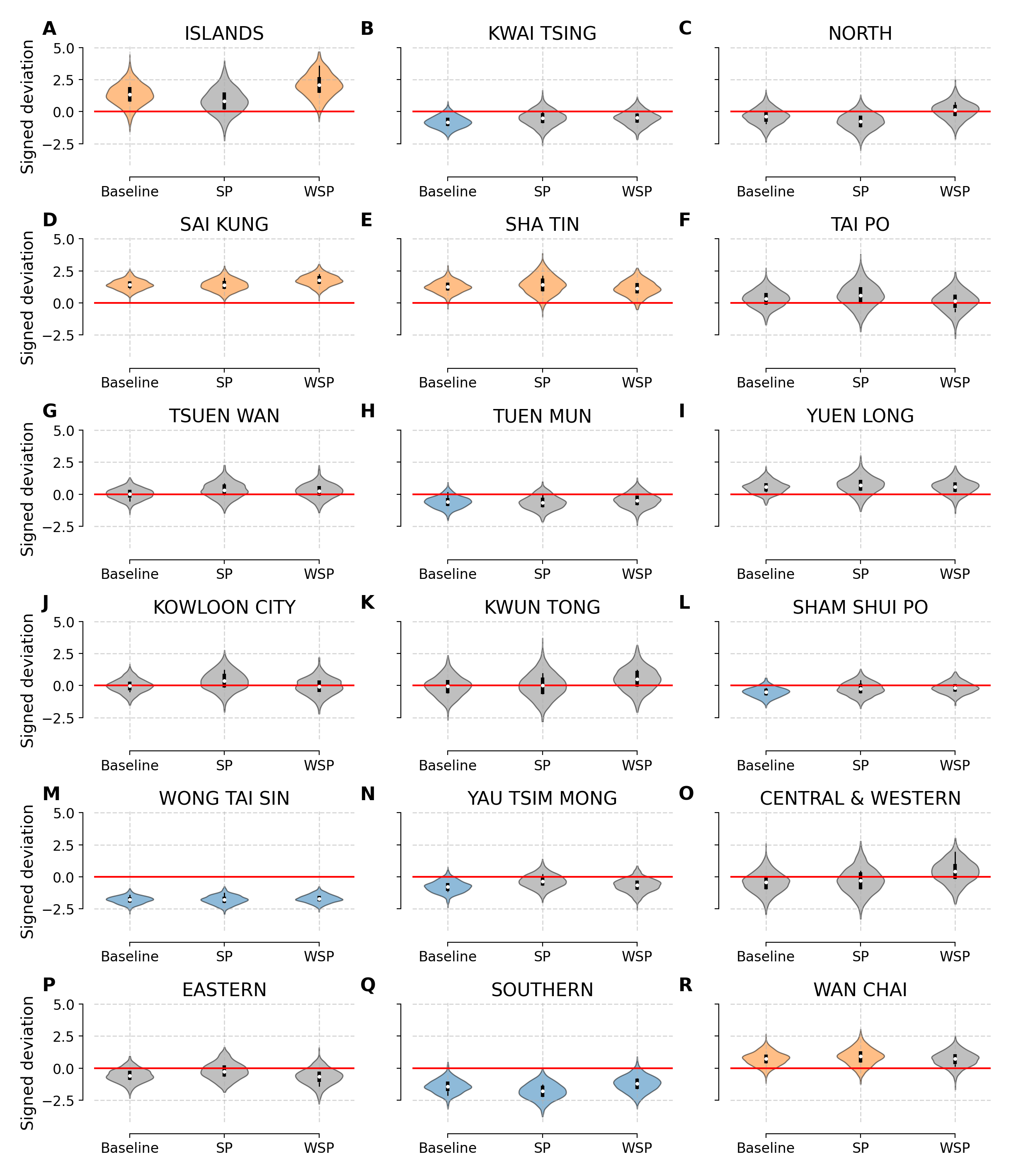}
\caption{ 
    \textbf{Relative likelihood of systematic bias for models trained on the default set of features in 2016}.
    We examine the distribution of probable outcomes of signed deviation by computing the difference between our predictions $\hat{Y}_i^{2016}$ and the ground truth mortality rates $Y_i^{2016}$ for each district. 
    A perfect model would have a narrow distribution centred on 0 (solid red line going across). Positive values show overestimation, whereas negative values show an underestimation of mortality rates for each district. 
    We color models with significance systematic overestimation in orange, while use blue to highlight models with significance systematic underestimation as measured by the $80\%$ CI.
}
\label{fig:errors_base_2016}   
\end{figure*} 

In Fig.~\ref{fig:errors_base_2016}, we display the empirical distributions of signed deviation to examine the relative likelihood of systematic bias for models trained on the default set of features in 2016.
We assess significance of model coefficients using centered $Q\%$ credible intervals (CI). 
A centered $Q$\% credible interval of a probability density function $p(x)$ is an interval $(a, b)$ defined such that $\frac{1}{2}(1 - \frac{Q}{100})= \int_{-\infty}^a\ dx\ p(x) = \int_b^\infty dx\ p(x)$.
We measure the significance of systematic errors in each model by computing the 80\% CI, whereby systematic overestimation is highlighted in orange ($\text{CI} > 0$), and systematic underestimation is colored in blue ($\text{CI} < 0$).

We note that our spatial models, especially the weighted spatial model, are effective at reducing systematic over- and under-estimations.
For example, the spatial models reduce the margin of error in panels B, H, L, and R of Fig.~\ref{fig:errors_base_2016}.
By contrast, all three models either overshoot or undershoot mortality rates drastically in a few districts (see panels D, E, M, and Q in~Fig.~\ref{fig:errors_base_2016}).

Our models also provide evidence to suggest that there are significant relationships between socioeconomic variables, such as household unemployment, percentage of single parents, and mortality rate. 
Many of these relations are significant in each of the census periods under study (2006, 2011, and 2016) while other relations are significant for some census periods but not others.
We display the distributions of $\beta$s in Fig.~\ref{fig:model_households_base_beta}---the parameter for the baseline model. 
For each panel, we show the kernel density estimation of $\beta$ as a function of each variable in the design tensor. 
We highlight distributions that are significantly above 0 using an orange color, while distributions significantly below 0 are colored in blue as measured by the $80\%$ CI.
Similar demonstrations of spatial and weighted spatial models can be found in Fig.~\ref{fig:spatial_model_households_base_beta}, and Fig.~\ref{fig:weighted_spatial_model_households_base_beta} respectively.
Besides the distributions of $\beta$s, we also show the kernel density estimation of $\gamma$s---the hyperparameter used for the spatial competent in each model in Fig.~\ref{fig:weighted_spatial_model_households_base_gamma}.

\begin{figure*}[tp!] 
\centering 
\includegraphics[width=\textwidth]{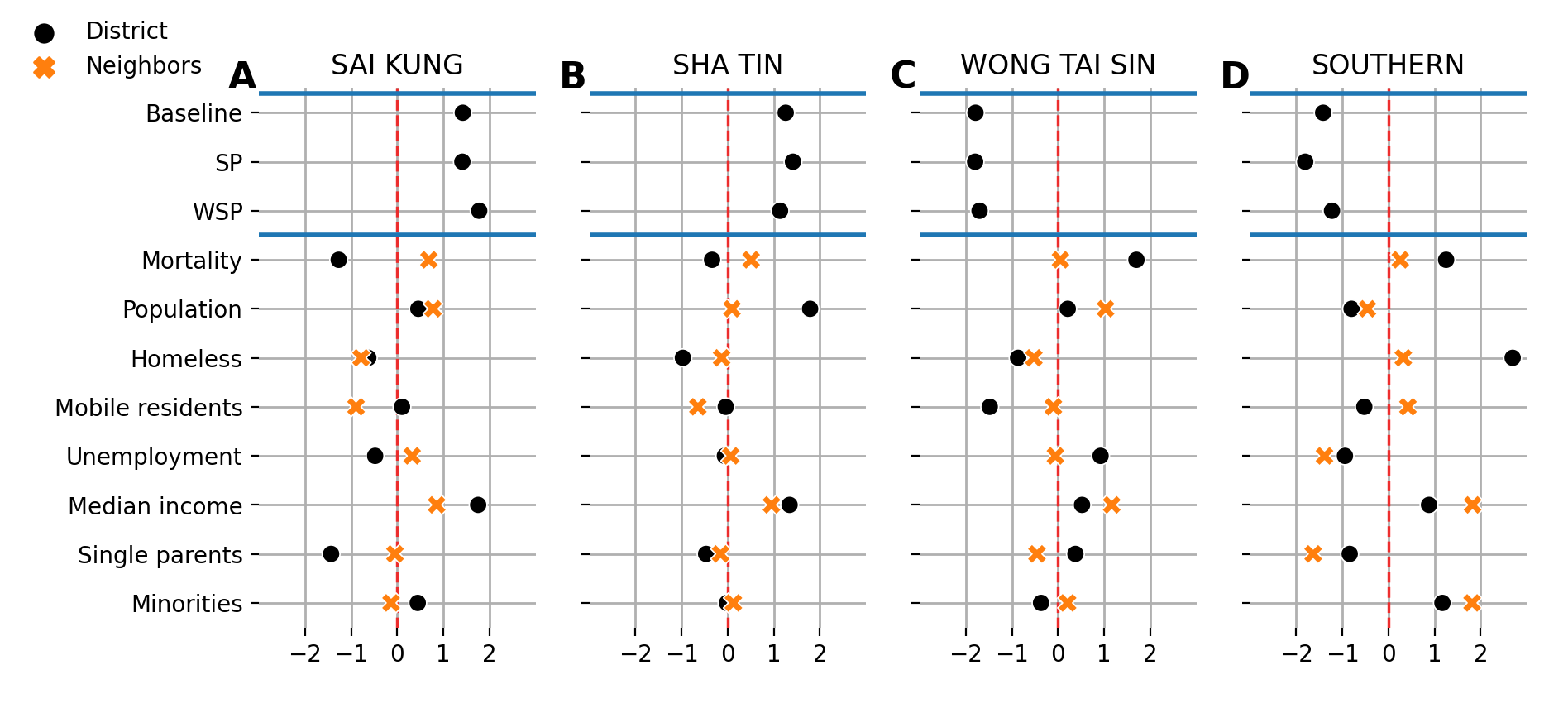} 
\caption{  
    \textbf{Impact of sociospatial factors on mortality risks.}.
    The first three rows show the mean signed deviation for four districts that are poorly fit by our models. 
    (\textbf{A--B}) We show districts with systematic overestimation of mortality rates, while (\textbf{C--D}) show districts where mortality rates are systematically underestimated. 
    For each district, we show the normalized value of some features of interest (black markers) along with the average value of the same features in the neighboring districts (orange markers).
    The red dashed line shows the average value for each of these normalized features centered at zero.
    We can see evidence of sociospatial factors of longevity in all four districts.
    Particularly, we note a spillover of wealth measured by median income.
    Districts in (\textbf{A}) and (\textbf{B}) maintain lower mortality rates while surrounded by districts with average mortality rates. 
    Districts in (\textbf{C}) and (\textbf{D}) have a socioeconomic pull, driving the entire neighborhood to have higher mortality rates.
} 
\label{fig:doi_base_2016}
\end{figure*}

All three models are fairly accurate nonetheless.
Access to a wider range of predictors and longitudinal data will help reduce our margin of error in estimating mortality rates.  
However, the spatial models allow us to capture nonlocal and interdependent interactions among the social and economic features across districts that would not be possible otherwise using the baseline model.

\subsection{Evidence of sociospatial spillovers}

\begin{figure*}[tp!] 
\centering 
\includegraphics[width=\textwidth]{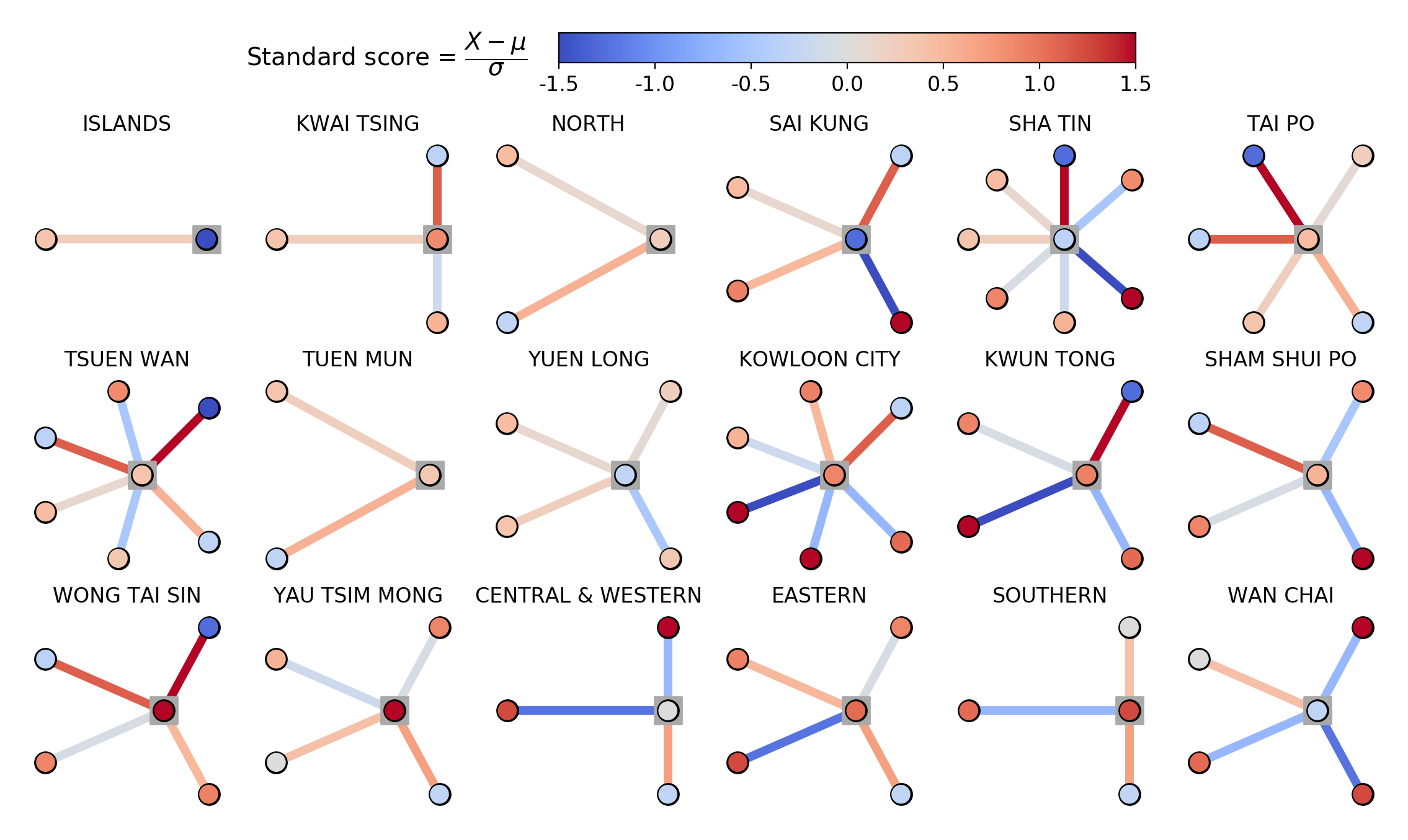} 
\caption{  
\textbf{Ego networks of each district demonstrating sociospatial factors of mortality for the 2016 weighted spatial model}.
We display ego networks of each district in Hong Kong and its nearest neighbors in the road and bridge network.  
The central node (highlighted with a grey box) of each network corresponds to the labelled district.
Neighbors are not arranged around the ego district geographically.
Node color corresponds to normalized mortality rate and edge color 
corresponds to signed prediction error for the 2016 WSP model.
These ego networks encode a qualitative measure of the sociospatial factors in mortality modeling.
We display the equivalent networks for the Baseline and SP models
in Fig.~\ref{fig:ego_base_2016_Baseline} and Fig.~\ref{fig:ego_base_2016_SP} respectively.
} 
\label{fig:ego_base_2016_WSP}
\end{figure*}  

The districts of Sai Kung, Sha Tin, Wong Tai Sin, and Southern are poorly fit by our models in 2016. 
In Fig.~\ref{fig:doi_base_2016}, we inspect the characteristics of each district and its neighbors. 
The first three rows show how all three models overestimate mortality rates for the Sai Kung and Sha Tin districts (Fig.~\ref{fig:doi_base_2016}A--B), while underestimating the other two districts (Fig.~\ref{fig:doi_base_2016}C--D). 
For each district, we plot the standardized score of some socioeconomic features (black markers).
We also display the corresponding average value of the same features from the neighboring districts derived from our network and denoted with orange markers to examine the effect of neighborhoods on these areas.

The first two districts (Fig.~\ref{fig:doi_base_2016}A--B) have lower mortality rates than average mortality of Hong Kong. They are significantly overestimated by our models, while mortality rates in their neighboring districts hover around average mortality rate of Hong Kong. 
The other two districts (Fig.~\ref{fig:doi_base_2016}C--D) have higher mortality rates and surrounded by districts with lower mortality rates.
The qualitative difference among these districts provides additional evidence of sociospatial and economic spillovers in mortality risks.
We can see spillovers of wealth in Figs.~\ref{fig:doi_base_2016}A--B with a higher median income in the district being associated with a higher median income in its neighborhood. 
Though districts in Figs.~\ref{fig:doi_base_2016}C--D are located in a wealthy neighborhood, they still have a lower median income. 
The Wong Tai Sin district has a lower number of mobile residents than average (see Fig.~\ref{fig:doi_base_2016}C). 
In Fig.~\ref{fig:doi_base_2016}D, we see an extraordinary higher number of homeless people compared to the neighboring population.
A higher percentage of minorities can also drive our systematic errors in this district, which hints at disparities in mortality risks in the area. 
We would need further analyses with finer geospatial resolution to explain this behavior.

We also analyze our signed deviation distributions through a local ego network approach. 
An ego network of a node is the network comprising that node and its nearest neighbors. 
Here each node is a district and its neighbors are that district’s neighboring districts. 
The joint distribution of mortality at time $t$ and district $i$ and the model mortality rate prediction conditioned on location at node $i$ can be concisely represented in the form of an ego network for each district. 
We display an example of this representation in Fig.\ \ref{fig:ego_base_2016_WSP}.
We color nodes by standardized mortality rate and edges by standardized signed deviation error. 
Though this is an exploratory method that deserves greater expansion and attention in future work, we note qualitatively that the local view of predicted mortality versus true mortality varies substantially as a function of district. 
For example, the district of Wan Chai is connected to four other districts, three of which have substantially higher mortality than average and in particular higher mortality than Wan Chai itself, which has lower mortality than average. 
The model predictions for these neighbors are lower than their true mortality. 
An observer in Wan Chai who has access only to the model predictions and not the true mortality data would rationally assume that mortality in these districts is much lower than it truly is and could subsequently make further inferences or decisions based on faulty-but-rational assumption.
Notably, socioeconomic diversity or heterogeneity of a neighborhood can change the local perception of mortality models across neighboring communities.
We observe that high divergence from the neighborhood is associated with higher rates of uncertainty in the models.  
We envision future work incorporating this sort of information to fine tune mortality models. 
 
\section{Concluding Remarks}\label{sec:conclusion}

Data-driven models are powerful tools often used to inform and reshape cultural, political, and financial policies around the globe. 
However, data scarcity and data sparsity pose an enormous challenge for some domains such as mortality modeling, especially for small territories. 
In this work, we studied the implications of that on the development of mortality models in Hong Kong with restricted access to publicly available data sources. 
We carried out a set of experiments to identify and explore how nonlocal and sociospatial interactions can systemically influence the outcome of a mortality model.

Our results support our hypothesis that spatial associations of wealth or social deprivation among neighborhoods have a direct and sometimes substantial impact on mortality risks.
Our examination reveals that localized models---which do not account for sociospatial factors---can systematically over- or under-estimate mortality rate---while spatial models reduce the error of predicting mortality rate.
In our investigation, we show how our models scale differently regarding their complexity and statistical inference.
We illustrate how the local perception of predicted mortality varies qualitatively and substantially as a function of the spatial unit. 
Future work can also improve upon our exploratory method to study spatial interdependency of social and economic factors of longevity, and identify sociospatial spillovers across neighbourhoods and communities.

We acknowledge our findings are limited for a few reasons. 
We only have access to census data for three individual years spanning a decade and a half. 
A better explanation of the nonlocalized effect of neighborhoods could be achieved by testing our hypotheses on additional years, along with more socioeconomic features to enrich our design tensor.

We used variational inference to estimate the posteriors analytically because of its effectiveness and versatility.
Although our decision of using variational Bayes is substantive, future studies can further explore and examine the costs and benefits for accurate estimation of posteriors using variational inference compared with other classical Bayesian methods.
Varying the model parameters temporally ensures a modular design, as access to richer and longitudinal sociotechnical data features continues to evolve.
However, our evaluation suggests that using fixed parameters over time can reduce the number of tuneable parameters in the models substantially to overcome the challenge of high bias-variance tradeoff of the spatially rich and larger models. 

Our geospatial resolution is unfortunately not high enough to identify some dynamics of connected communities.  
Our method, however, can be implemented similarly regardless of the spatial unit used for the experiment. 
Our spatial network is mainly based on the road network of Hong Kong, which could be extended to account for roads/bridges and public transport across any desired spatial unit.

Finally, we have only explored a distance-based weighting scheme for the connections across districts in the network. 
Population density could be included to enrich the socioeconomic effect of neighboring regions on a node within the network (for example, theory of intervening opportunities~\cite{stouffer1940intervening}). 
Other attributes such as geographic information associated with community health services could help us assess their value and reallocate these centers to more optimized locations.

\begin{acknowledgments}
The authors are grateful for the computing resources provided by the Vermont Advanced Computing Core 
and financial support from the Massachusetts Mutual Life Insurance Company and Google Open Source under the Open-Source Complex Ecosystems And Networks (OCEAN) project.
While they did not collaborate on the manuscript, 
we thank Hong Kong's Census and Statistics Department for facilitating access to their mortality dataset.
The authors are grateful for useful conversations with Adam Fox, Marc Maier, and Xiangdong Gu.
We also thank Josh Minot, Michael Arnold, Anne Marie Stupinski, Colin Van Oort, 
and many of our colleagues at the Computational Story Lab for their discussions and feedback on this project. 
\end{acknowledgments}

\bibliography{references}
%%%%%%%%%%%%%%%%%%%%%%%%%%%%%%%%%%%%%%%%%%%%%%%%%%%%%%%%%%%

%% supplementary

%% following records the starting page number of the supplementary
%% section in a file called startsupp.txt
%% enables script-based breaking of manuscript and supplementary
\newwrite\tempfile
\immediate\openout\tempfile=startsupp.txt
\immediate\write\tempfile{\thepage}
\immediate\closeout\tempfile

\renewcommand{\thefigure}{S\arabic{figure}}
\renewcommand{\thetable}{S\arabic{table}}
\setcounter{figure}{0}
\setcounter{table}{0}

\onecolumngrid
\appendix
\newpage

\section{Data variables}\label{sec:data_variables}

\begin{figure*}[h!]
\centering
\includegraphics[width=\textwidth]{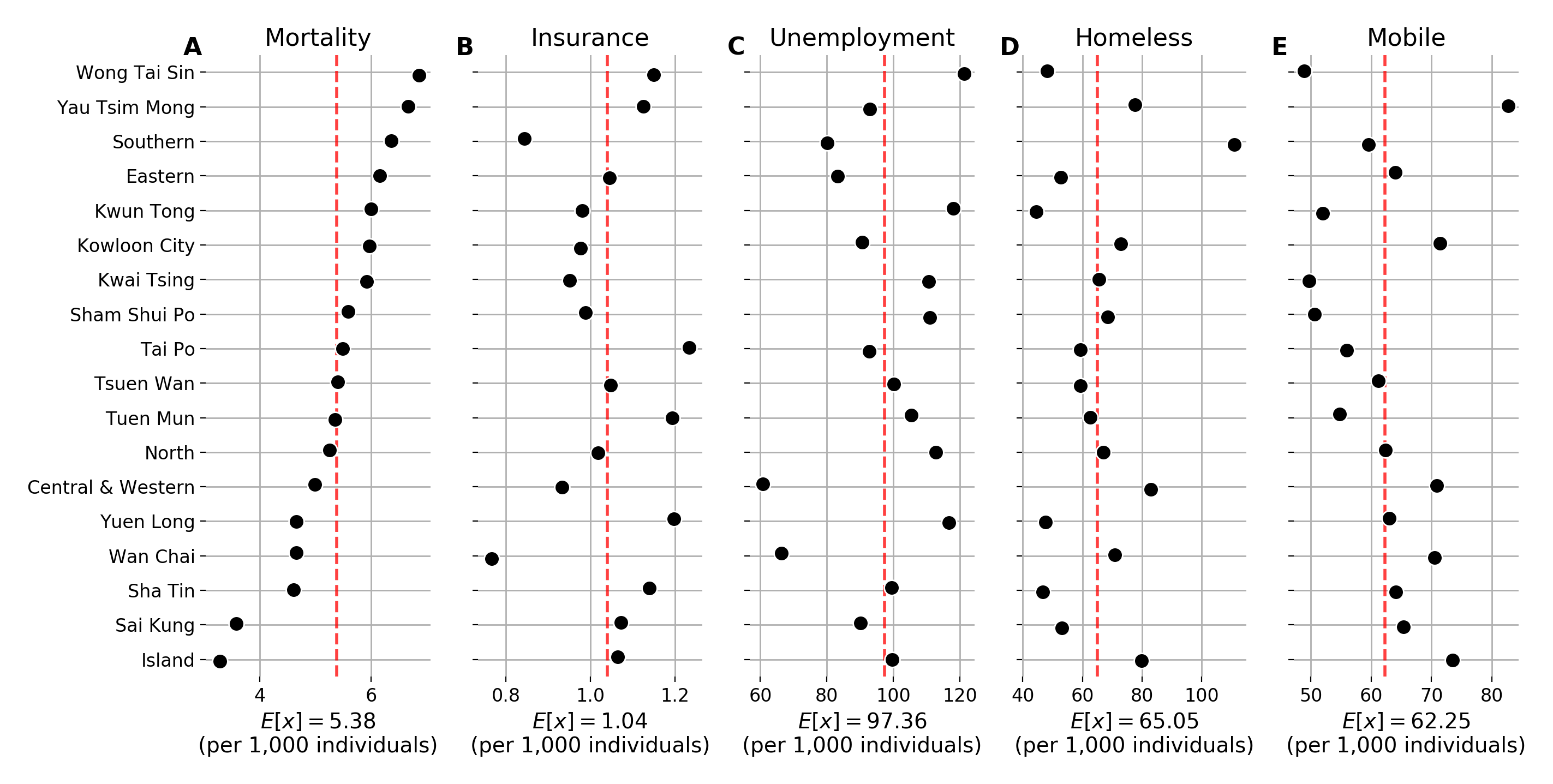}
\caption{
    \textbf{Cross-sectional attributes for calendar year 2016 by district} including
    \textbf{(A)} mortality rates,
    \textbf{(B)} number of life insurance policies,
    \textbf{(C)} unemployment rates,
    \textbf{(D)} number of homeless individuals, and
    \textbf{(E)} number of mobile residents.
    The dashed red lines indicate the average value for each of these variables in Hong Kong. 
    The four attributes presented here are not strongly correlated with mortality.
}
\label{fig:attrs_2016}
\end{figure*} 
 
\begin{figure*}[tp!]
\centering
\includegraphics[width=\textwidth]{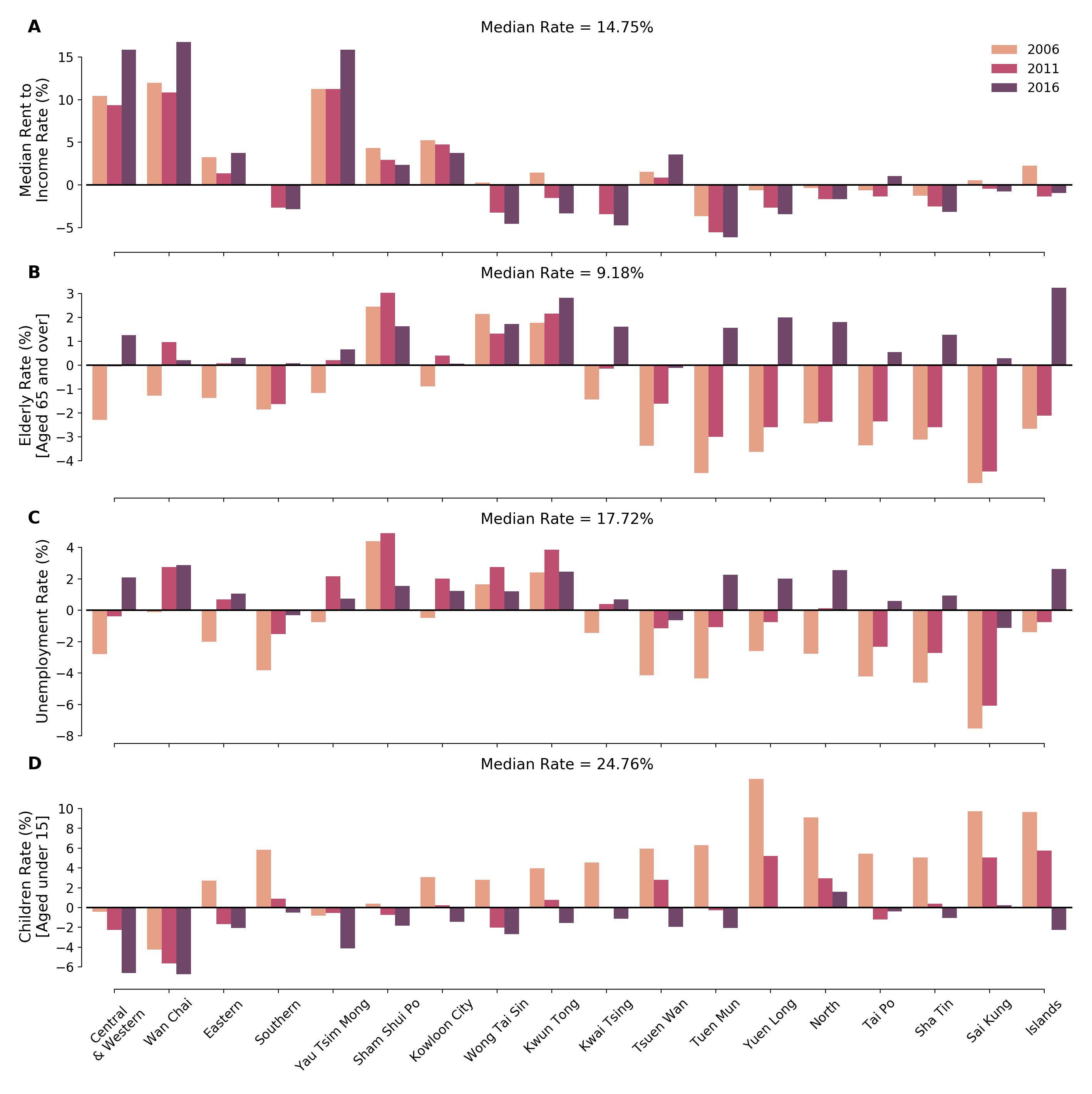}
\caption{
    \textbf{Households statistics by year.} 
}
\label{fig:household}
\end{figure*}

\begin{itemize}
\item \textbf{Mortality rates}:
Annual death rates by districts provided by the Census and Statistics Department of Hong Kong~\cite{hongkong_mortality_dataset}. 
We derive an annual crude death rate of each district by simply dividing the number of registered death records by the population size for each calendar year such that: 
$$Y_{it} = \dfrac{N_{it}(\textnormal{deaths})}{N_{it}(\textnormal{census population})},$$
for district $i$ and $t \in [2006, 2011, 2016]$.
\item \textbf{MMA life insurance coverage}:
Annual proportion of insurance policies issued in each district and normalized by population size.  
\item \textbf{Median income}:
Estimation of median income by the Census and Statistics Department of Hong Kong for each district. 
\item \textbf{Median rent to income ratio}:
The percentage of monthly household income spent on monthly household rent in inflation-adjusted terms. 
\item \textbf{Unemployment rates}:
The proportion of people without an active job normalized by population size. 
This is a proxy variable derived from the gap between the available labour force and
economically active individuals---aged over 15 and has been working for at least a week in the reference year~\cite{hongkong_census_dataset}. 
\item \textbf{Proportion of children}:
The proportion of young children aged under 15 for each district.
\item \textbf{Proportion of elderly}:
The proportion of seniors aged 65 or above for each district.
\item \textbf{Median monthly household income}:
Similar to median income, but this is computed for households rather than individuals.  
\item \textbf{Unemployment rates across households}:
Similar to unemployment rates but derived at the households level rather than individuals. 
\item \textbf{Unemployment rates among minorities}:
Similar to unemployment rates but computed for minorities only. 
\item \textbf{Proportion of homeless people}:
Another proxy variable to estimate the proportion of people not living in registered domestic households. 
\item \textbf{Proportion of homeless mobile residents}: 
Hong Kong's permanent residents 
who had stayed in a given district for ``at least 1 month but less than 3 months'' 
during the the 6 months before or after the reference moment~\cite{hongkong_census_dataset}.
\item \textbf{Proportion of single parents}:
``Mothers or fathers who are never married,
widowed, divorced or separated, with
child(ren) aged under 18 living with them in
the same household''~\cite{hongkong_census_dataset}.
\item \textbf{Proportion of households with children in school}:
The proportion of households with young adults attending full-time courses in educational institutions in Hong Kong.
\end{itemize}

\begin{figure*}[tp!]
\centering
\includegraphics[width=\textwidth]{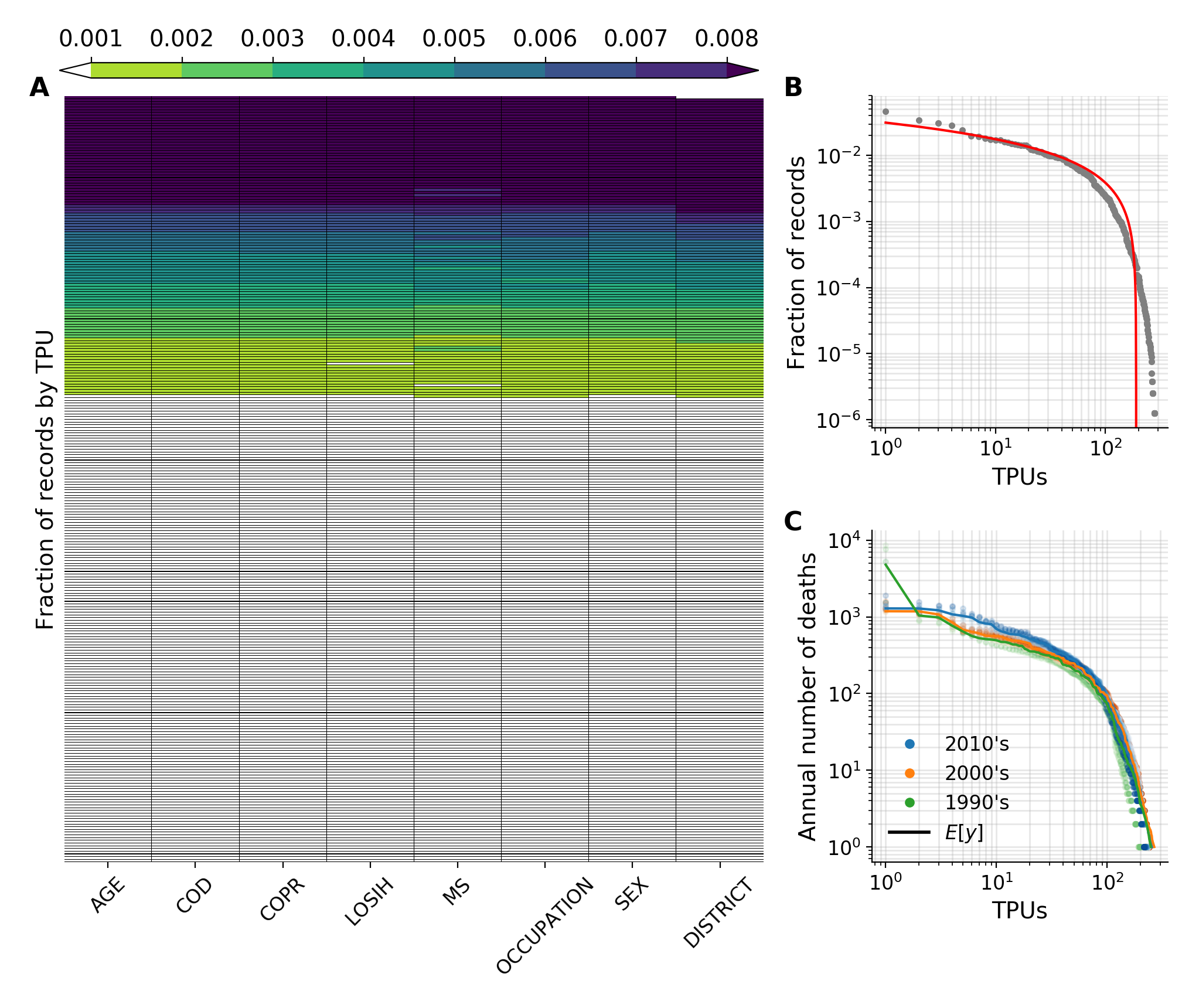}
\caption{
    \textbf{Records by TPU.} 
    \textbf{(A)} shows the fraction of records reported by each TPU in the mortality dataset.
    \textbf{(B)} shows an exponential declines in the number of records by TPU
    whereby the horizontal axis shows TPUs ranked by the fraction of death records 
    reported on the vertical axis. Similarly,  
    \textbf{(C)} shows the annual average of deaths for each decade by TPU. 
}
\label{fig:tpus}
\end{figure*} 

Our initial analysis shows variations in mortality rates across districts. 
In Fig.~\ref{fig:attrs_2016}, we show a cross-sectional analysis of 4 different socioeconomic indices with respect to our response variable (mortality rate) for 2016. 
We normalize these values to show the number of people in each bin per 1,000 individuals for each variable. 
The dashed red lines show the average value. 
Districts with a lower number of mobile residents have higher rates of morality, such as Wong Tai Sin, Kwun Tong, Kwai Tsing, and Sham Shui Po. 
We note high mortality rates in the Yau Tsim Mong and Southern districts, while having a higher number of homeless people. 
Many districts with a lower number of homeless people have higher rates of mortality, such as Wong Tai Sin, Eastern, and Kwun Tong, and Kowloon City.

\section{Tertiary Planning Unit (TPU)}\label{sec:tpu}

\begin{figure*}[h!]
\centering
\includegraphics[width=\textwidth]{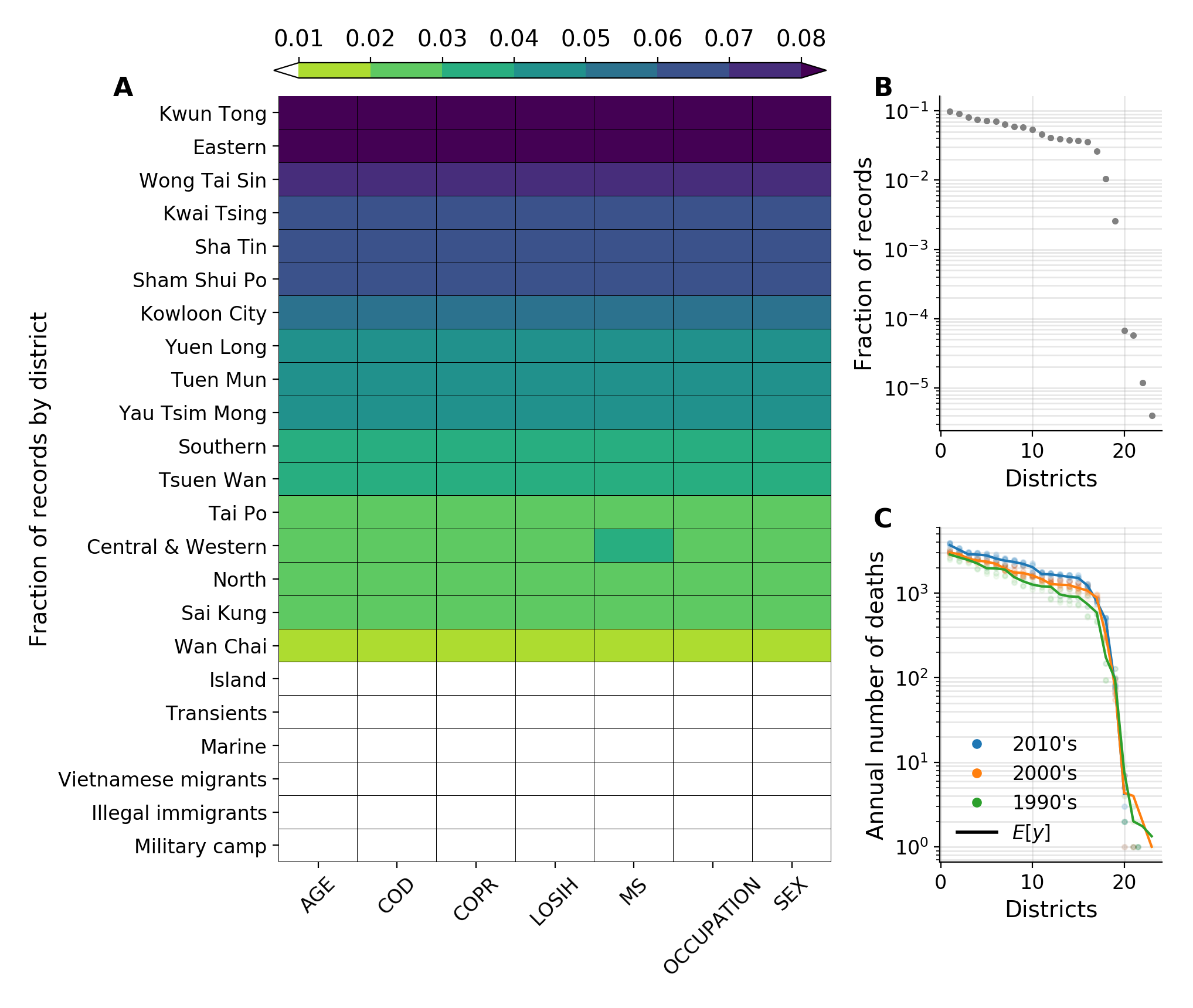}
\caption{
    \textbf{Records by district.} 
    \textbf{(A)} shows the fraction of records reported by each district in the mortality dataset.
    \textbf{(B)} shows an exponential declines in the number of records by district
    whereby the horizontal axis shows districts ranked by the fraction of death records 
    reported on the vertical axis. Similarly, 
    \textbf{(C)} shows the annual average of deaths for each decade by district. 
}
\label{fig:districts}
\end{figure*}

\begin{figure*}[tp!]
\centering
\includegraphics[width=\textwidth]{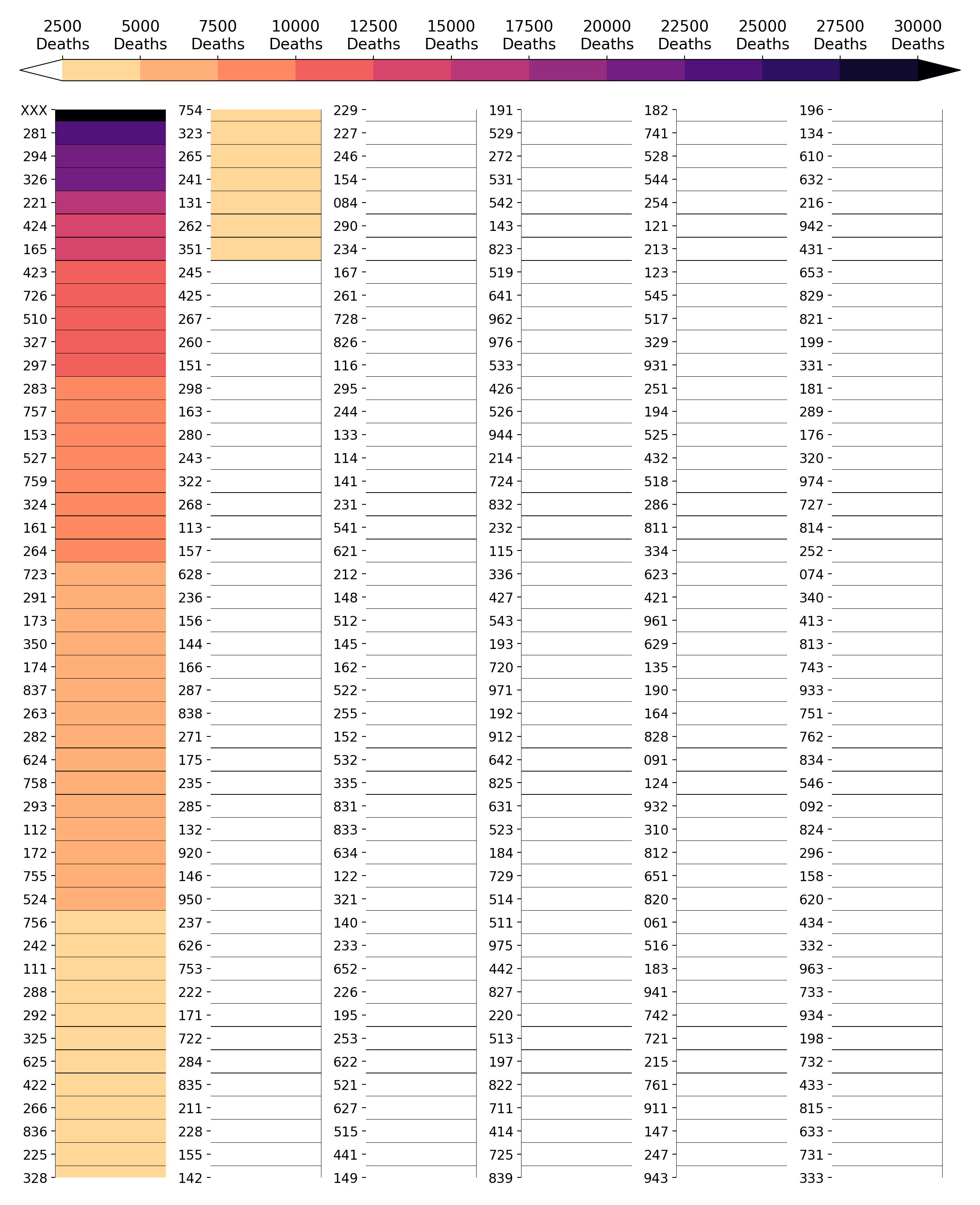}
\caption{
    \textbf{Annual number of deaths by TPU.} 
    Death records are not reported frequently
    at the TPU level in our mortality dataset. 
    We note that the code (XXX) is used to indicate missing labels.
}
\label{fig:deaths_tpus}
\end{figure*}

A Tertiary Planning Unit (TPU)~\cite{tpu}---similar to a census-block in the US---is a geospatial reference system used by the 
Hong Kong’s Census and Statistics Department to report population census statistics.
To inspect the geospatial tags we have in our dataset, we show the fraction of death records found in our data set by each TPU (Fig.~\ref{fig:tpus}) and by districts (Fig.~\ref{fig:districts}).
Unfortunately, most death records were found in a small fraction of TPUs, which is understandable because of known data privacy concerns about the dataset, as records in small TPUs may reveal sensitive information about specific individuals there.
In Fig.~\ref{fig:deaths_tpus}, we show the annual number of reported deaths by TPU.
We note that death records are not reported frequently at the TPU level in the dataset.
Missing records are flagged with XXX as stated in the data set dictionary provided by the Hong Kong’s Census and Statistics Department.

To avoid any risk of identifying individuals in the dataset, we use districts as our main spatial unit, trying our best to stay away from the pitfalls of data privacy in census data. 
In fact, most studies have either filtered out small TPUs in their analyses~\cite{wong2008effects,chung2018socioeconomic}, or aggregated their records at the district level~\cite{thach2015assessing} to overcome this challenge in the dataset.
In Fig.~\ref{fig:districts}, the main 18 districts of Hong Kong show very consistent data distributions for all variables in the data set throughout the last few decades. 
We note that the last 5 entries in the heatmap are not districts per se, but they are only used by the census department to report deaths outside the borders of the main 18 districts.
However, given access to a more granular geospatial resolution (\textit{i.e.,} TPU) our method can be implemented in a similar manner to identify and explore some of these sociotechnical dynamics.

\section{Supplementary Figures}

\begin{figure*}[th!]
\centering
\includegraphics[width=\textwidth]{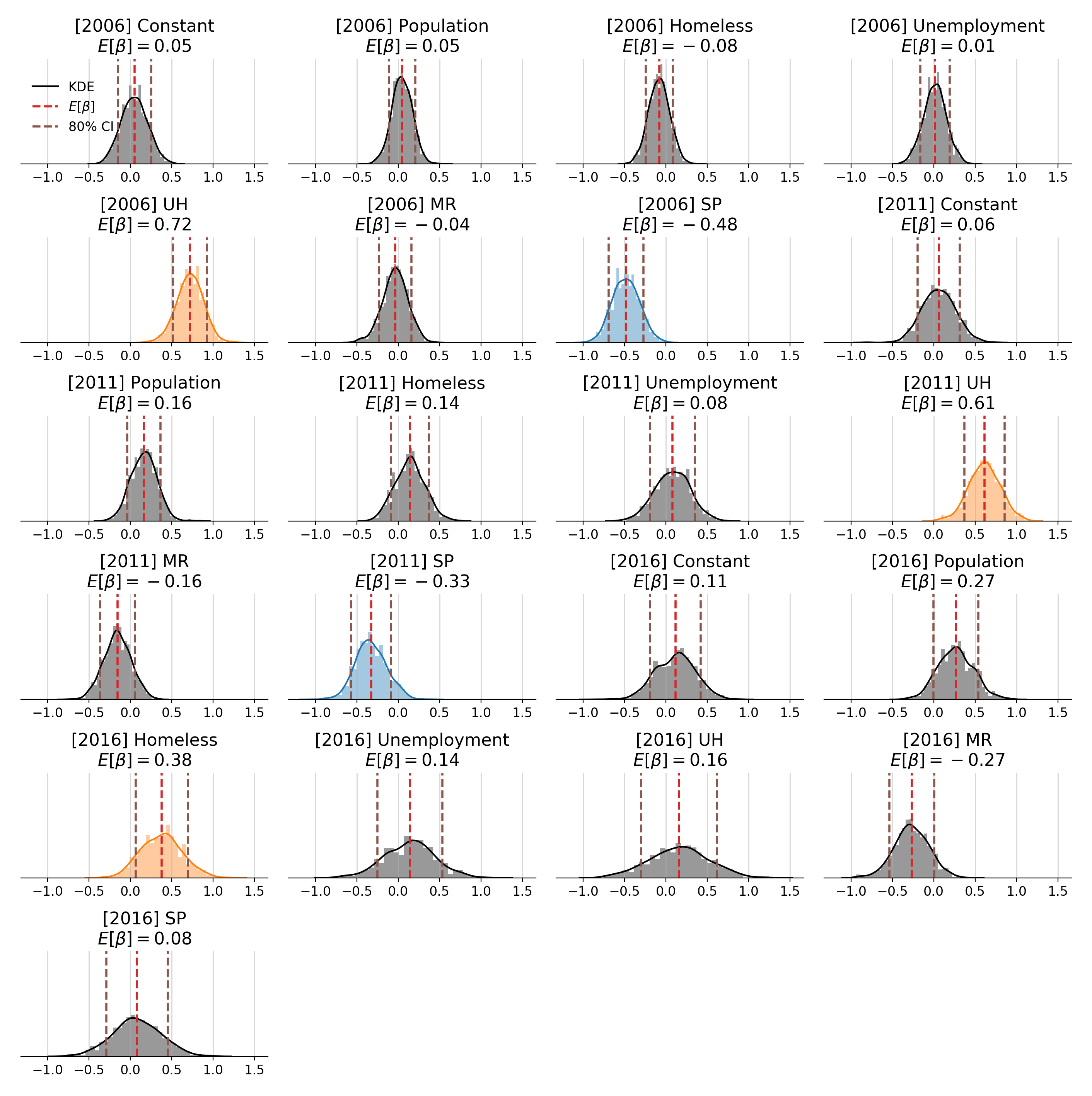}
\caption{
    \textbf{Distributions of $\beta$ for the baseline model.}
    We show the kernel density estimation of $\beta$ coefficient for each feature in the design tensor.  
    Distributions that are significantly above the mean are colored in orange
    while distributions significantly below the mean are colored in blue
    as measured by the $80\%$ CI.
}
\label{fig:model_households_base_beta}
\end{figure*}

\begin{figure*}[tp!]
\centering
\includegraphics[width=\textwidth]{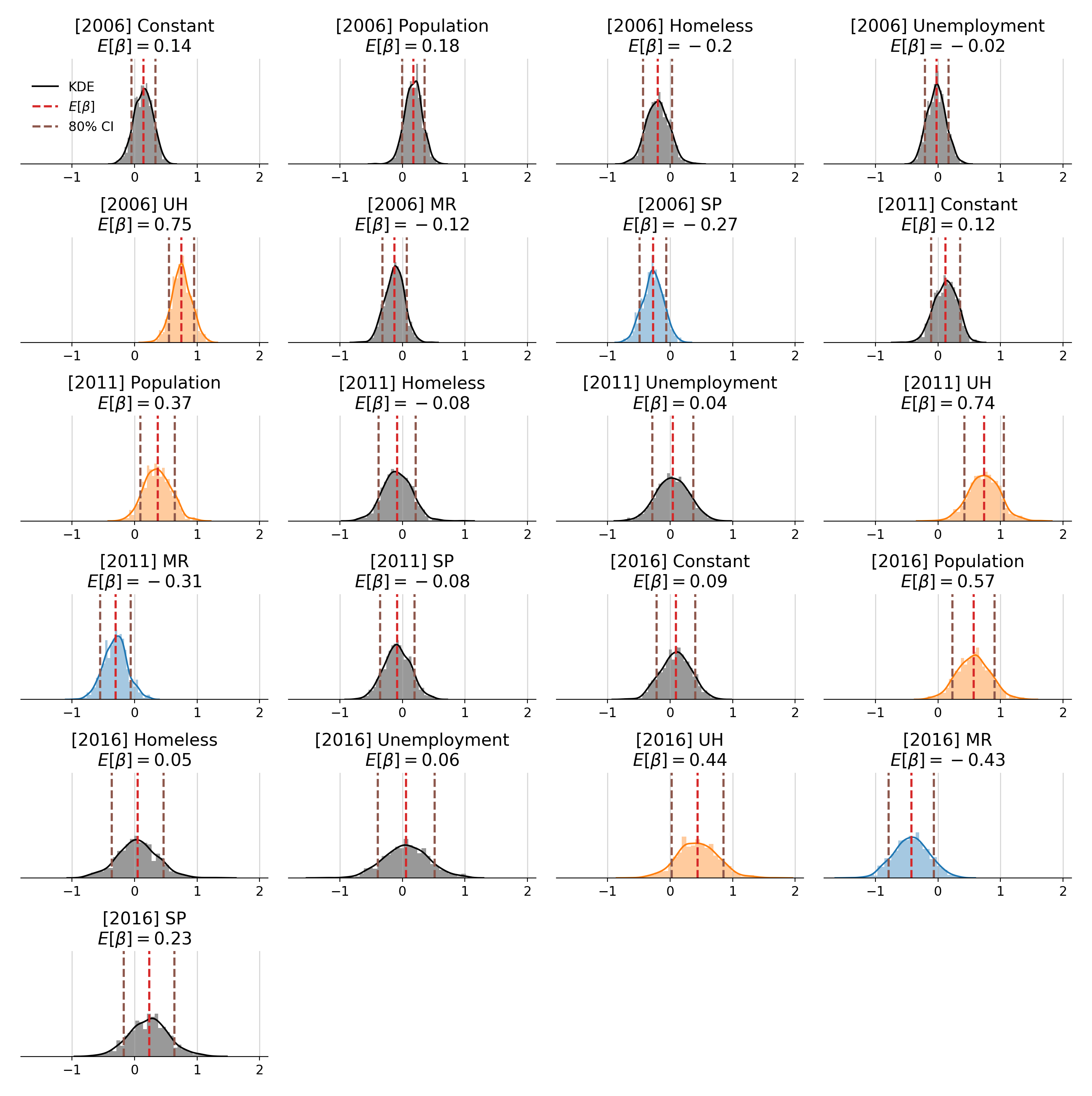}
\caption{
    \textbf{Distributions of $\beta$ for the spatial model.}
    We show the kernel density estimation of $\beta$ coefficient for each feature in the design tensor.  
    Distributions that are significantly above the mean are colored in orange
    while distributions significantly below the mean are colored in blue
    as measured by the $80\%$ CI.
}
\label{fig:spatial_model_households_base_beta}
\end{figure*}

\begin{figure*}[tp!]
\centering
\includegraphics[width=\textwidth]{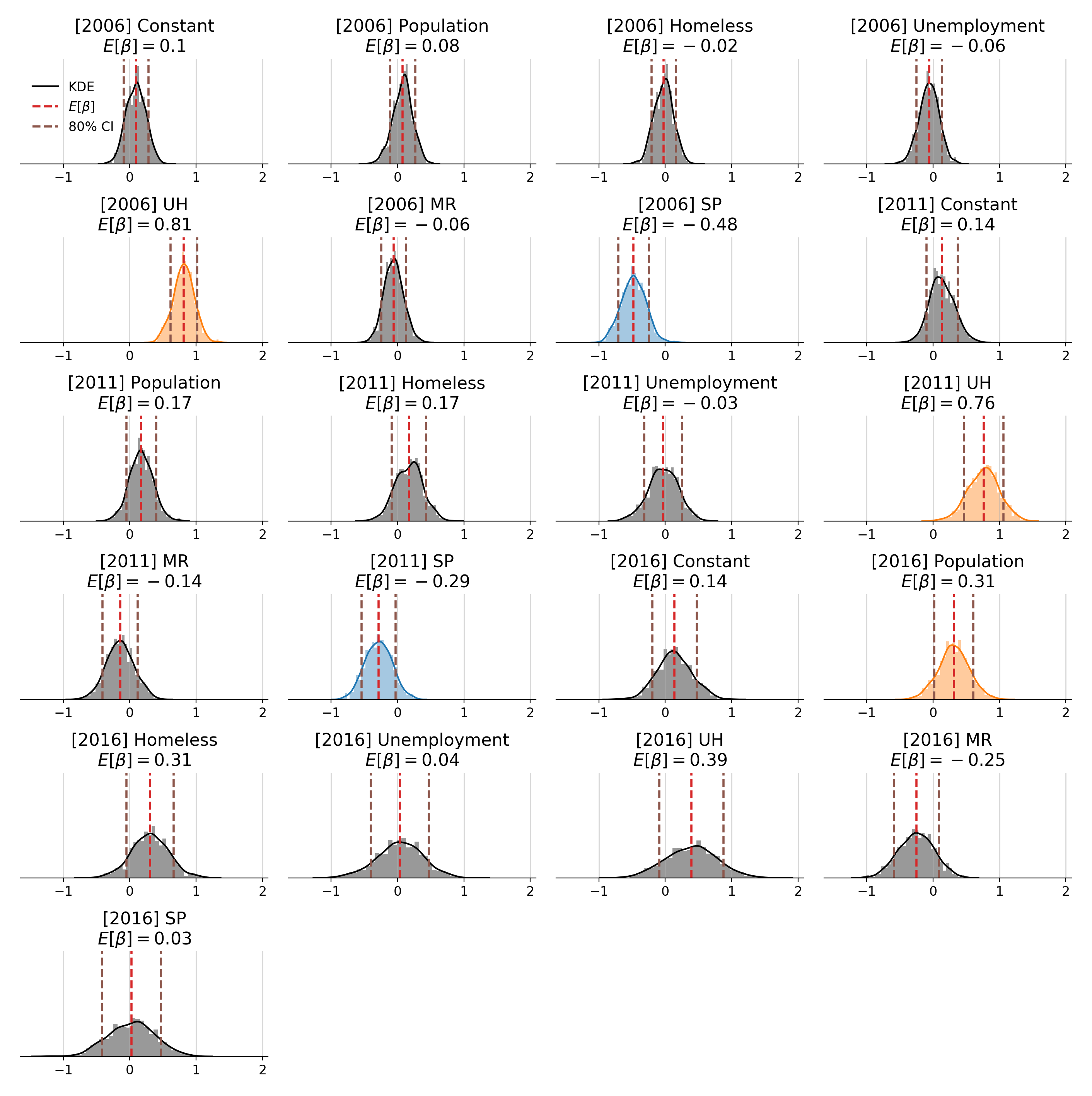}
\caption{
    \textbf{Distributions of $\beta$ for the weighted spatial model.}
    We show the kernel density estimation of $\beta$ coefficient for each feature in the design tensor.  
    Distributions that are significantly above the mean are colored in orange
    while distributions significantly below the mean are colored in blue
    as measured by the $80\%$ CI.
}
\label{fig:weighted_spatial_model_households_base_beta}
\end{figure*}

\begin{figure*}[tp!] 
\centering
\includegraphics[width=\textwidth]{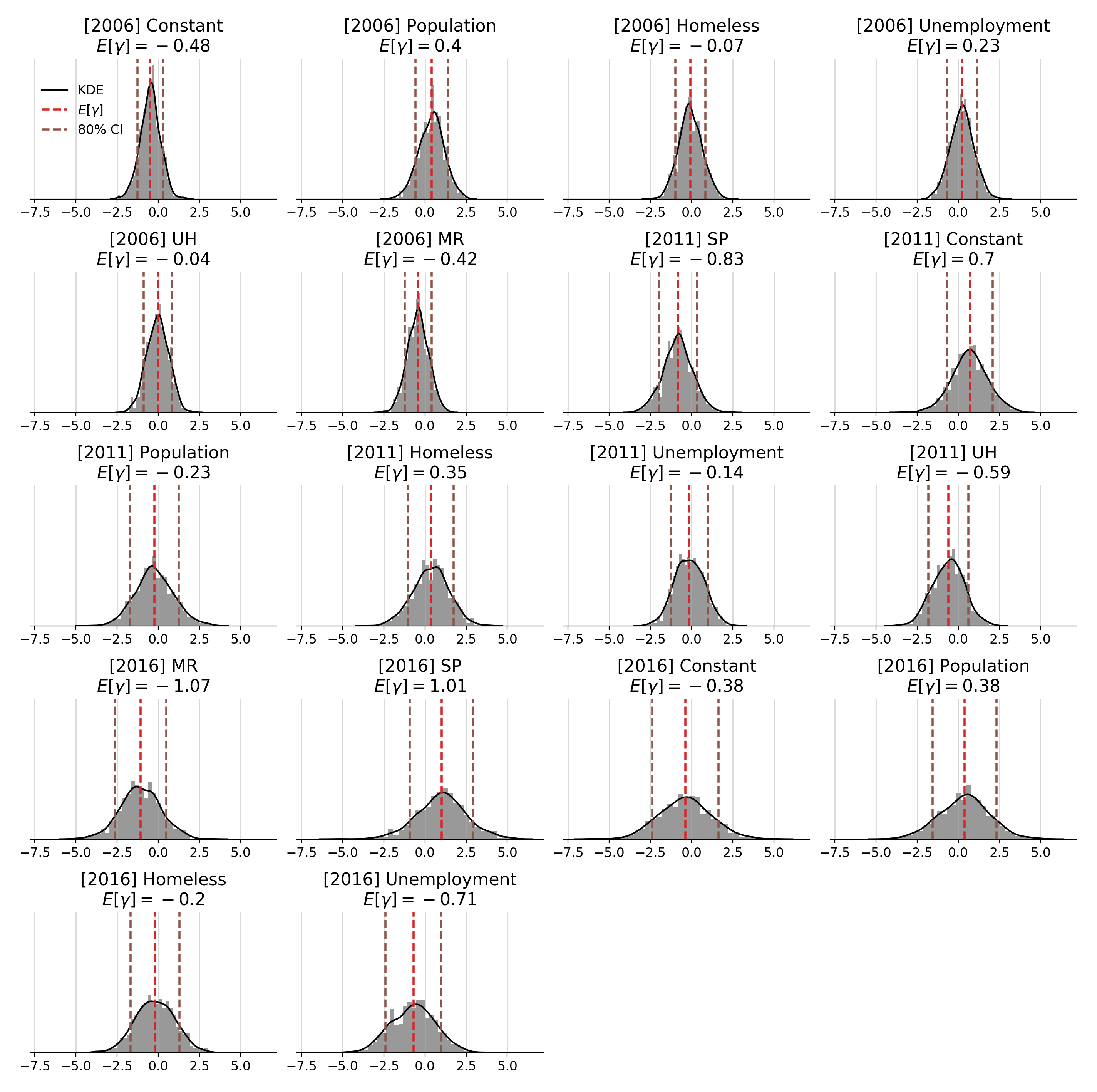}
\caption{
    \textbf{Distributions of $\gamma$ for the weighted spatial model.}
    We show the kernel density estimation of $\gamma$ coefficient for each feature in the design tensor.  
    Distributions that are significantly above the mean are colored in orange
    while distributions significantly below the mean are colored in blue
    as measured by the $80\%$ CI.
}
\label{fig:weighted_spatial_model_households_base_gamma}
\end{figure*}

\begin{figure*}[tp!] 
\centering 
\includegraphics[width=.8\textwidth]{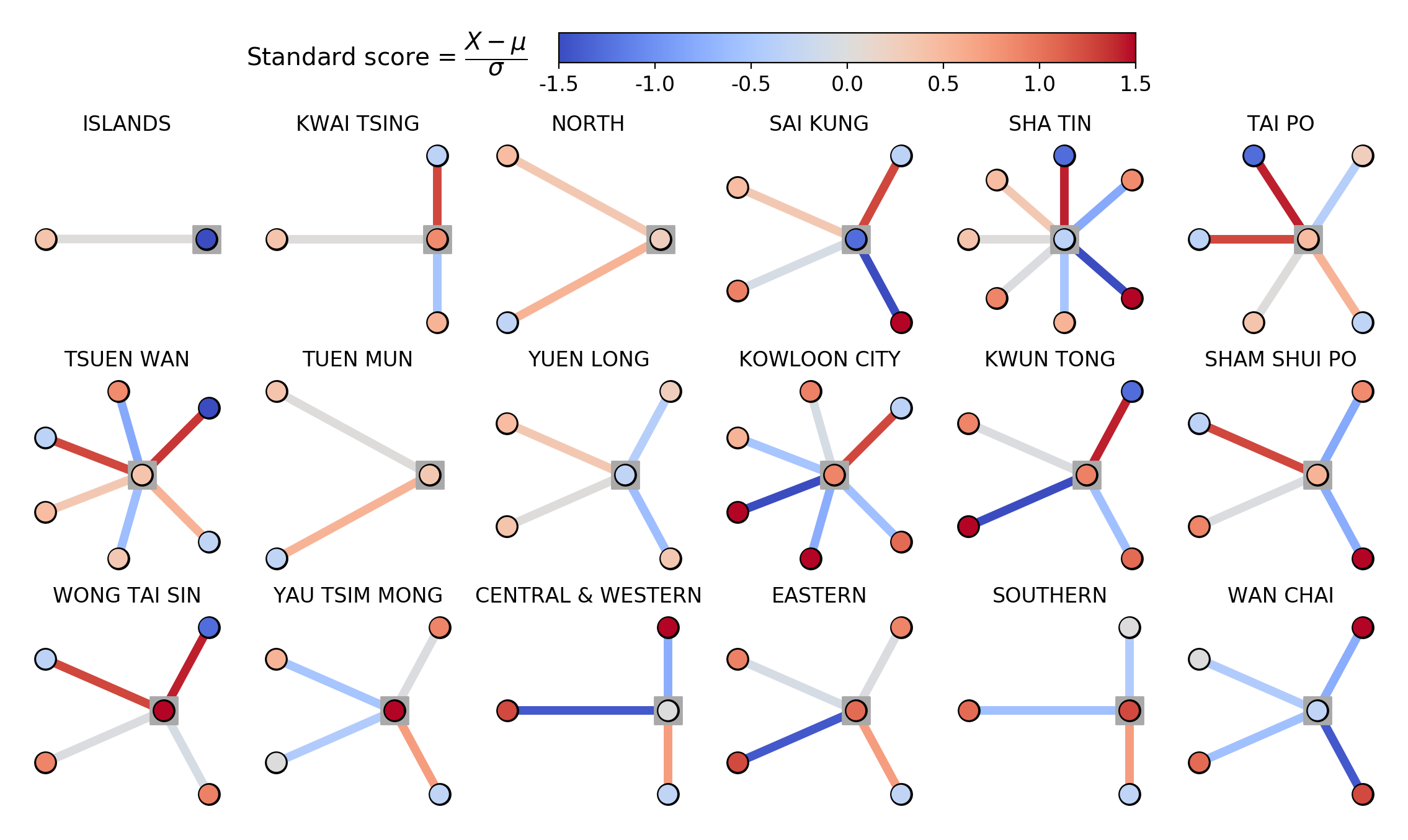} 
\caption{ 
\textbf{Ego networks of each district demonstrating sociospatial factors of mortality for the 2016 baseline model}.
We display ego networks of each district in Hong Kong and its nearest neighbors in the road and bridge network.  
The central node (highlighted with a grey box) of each network corresponds to the labelled district.
Neighbors are not arranged around the ego district geographically.
Node color corresponds to normalized mortality rate and edge color 
corresponds to signed prediction error for the 2016 WSP model.
These ego networks encode a qualitative measure of the sociospatial factors in mortality modeling.
}
\label{fig:ego_base_2016_Baseline}
\end{figure*}

\begin{figure*}[tp!] 
\centering 
\includegraphics[width=.8\textwidth]{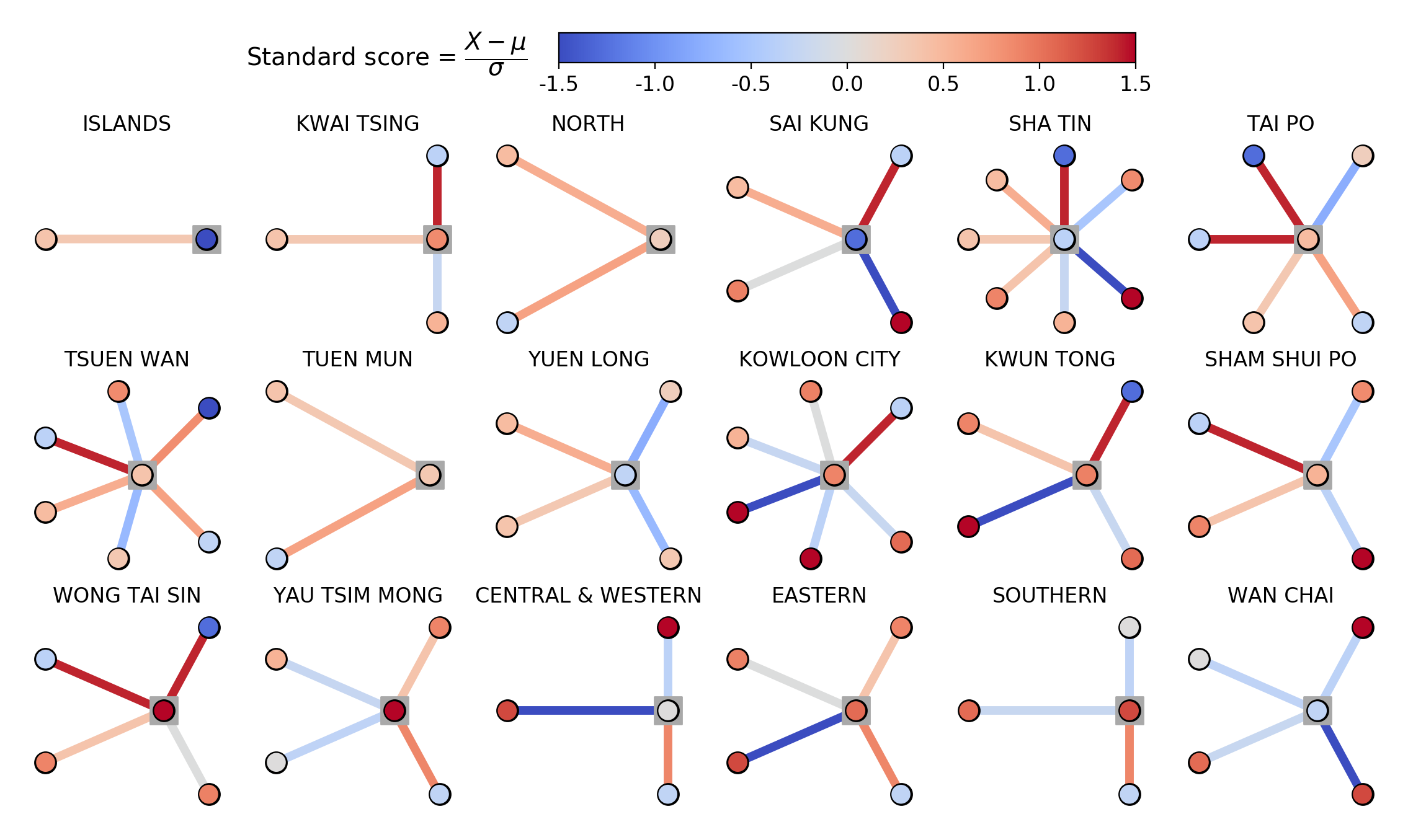} 
\caption{ 
\textbf{Ego networks of each district demonstrating sociospatial factors of mortality for the 2016 spatial model}.
We display ego networks of each district in Hong Kong and its nearest neighbors in the road and bridge network.  
The central node (highlighted with a grey box) of each network corresponds to the labelled district.
Neighbors are not arranged around the ego district geographically.
Node color corresponds to normalized mortality rate and edge color 
corresponds to signed prediction error for the 2016 WSP model.
These ego networks encode a qualitative measure of the sociospatial factors in mortality modeling.
}
\label{fig:ego_base_2016_SP}
\end{figure*}  

\end{document}